\documentclass[twocolumn]{aastex62}
\usepackage{amsmath,amstext}
\usepackage[T1]{fontenc}
\usepackage{apjfonts} 
\usepackage[figure,figure*]{hypcap}
\usepackage{natbib}
\usepackage[disable,colorinlistoftodos]{todonotes}

\usepackage{soul, CJK}

\let\Oldtodo\todo
\renewcommand{\todo}[1]{\Oldtodo[inline]{#1}}

\shorttitle{The Zwicky Transient Facility}
\shortauthors{Bellm et al.}

\begin{document}
\title{The Zwicky Transient Facility: System Overview, Performance, and First Results}

\begin{CJK*}{UTF8}{bkai}

\author[0000-0001-8018-5348]{Eric C. Bellm}
\affiliation{DIRAC Institute, Department of Astronomy, University of Washington, 3910 15th Avenue NE, Seattle, WA 98195, USA}

\author[0000-0001-5390-8563]{Shrinivas R. Kulkarni}
\affiliation{Division of Physics, Mathematics, and Astronomy, California Institute of Technology, Pasadena, CA 91125, USA}

\author[0000-0002-3168-0139]{Matthew J. Graham}
\affiliation{Division of Physics, Mathematics, and Astronomy, California Institute of Technology, Pasadena, CA 91125, USA}

\author{Richard Dekany}
\affiliation{Caltech Optical Observatories, California Institute of Technology, Pasadena, CA 91125, USA}

\author{Roger M. Smith}
\affiliation{Caltech Optical Observatories, California Institute of Technology, Pasadena, CA 91125, USA}

\author{Reed Riddle}
\affiliation{Caltech Optical Observatories, California Institute of Technology, Pasadena, CA 91125, USA} 

\author[0000-0002-8532-9395]{Frank J. Masci}
\affiliation{Infrared Processing and Analysis Center, California Institute of Technology, MS 100-22, Pasadena, CA 91125, USA}

\author[0000-0003-3367-3415]{George Helou}
\affiliation{Infrared Processing and Analysis Center, California Institute of Technology, MS 100-22, Pasadena, CA 91125, USA}

\author{Thomas A. Prince}
\affiliation{Division of Physics, Mathematics, and Astronomy, California Institute of Technology, Pasadena, CA 91125, USA}

\author{Scott M. Adams}
\affiliation{Division of Physics, Mathematics, and Astronomy, California Institute of Technology, Pasadena, CA 91125, USA}

% ...

\begin{abstract}

The Zwicky Transient Facility (ZTF) is a new optical time-domain survey that uses the Palomar 48-inch Schmidt telescope.
A custom-built wide-field camera provides a 47\,deg$^2$ field of view and 8\,second readout time, yielding more than an order of magnitude improvement in survey speed relative to its predecessor survey, the Palomar Transient Factory (PTF).
We describe the design and implementation of the camera and observing system.
The ZTF data system at the Infrared Processing and Analysis Center provides near-real-time reduction to identify moving and varying objects.
We outline the analysis pipelines, data products, and associated archive.
Finally, we present on-sky performance analysis and first scientific results from commissioning and the early survey.
ZTF's public alert stream will serve as a useful precursor for that of the Large Synoptic Survey Telescope.

\end{abstract}

\listoftodos


\section{Introduction}

Large optical sky surveys have served as engines of discovery throughout the history of astronomy.
By cataloging large samples of astrophysical objects, these surveys provide literal and metaphorical finder charts for detailed followup observations with larger and more expensive telescopes.

In the last century, among the most influential sky surveys were the Palomar Observatory Sky Surveys.  
Conducted with photographic plates using the wide-field Palomar 48-inch Schmidt telescope \citep{Harrington:52:P48}, the first and second sky surveys (POSS-I, \citealp{Minkowski:63:POSS-I}; POSS-II, \citealp{Reid:91:POSS-II})
mapped the Northern Hemisphere sky and enabled fifty years of discovery.
A digitized version\footnote{\url{http://stdatu.stsci.edu/dss/}} \citep{Lasker:94:PlateDigitization, Djorgovski:98:DPOSS} is still widely used today.

The advent of solid-state charge coupled devices (CCDs) provided a huge leap in the quantum efficiency (QE) of astronomical cameras, enabling existing telescopes to reach greater depths with shorter exposures.
Contemporaneous improvements in CCD controller readout time and computer processing speed have increased data volumes while allowing data processing to keep pace.
Beginning especially with the Sloan Digital Sky Survey \citep[SDSS;][]{York:00:SDSS} but also including the Optical Gravitational Lensing Experiment \citep[OGLE;][]{Udalski:92:OGLE}, the All-Sky Automated Survey \citep[ASAS;][]{Pojmanski:97:ASAS}, the Lincoln Near-Earth Asteroid Research survey \citep[LINEAR;][]{Stokes:00:LINEAR}, the Supernova Legacy Survey \citep[SNLS;][]{Astier:06:SNLSY1}, Palomar-Quest \citep{Djorgovski:08:PalomarQuest}, the Catalina Sky Survey \citep[CSS;][]{Larson:03:CSS} and associated Catalina Real-Time Transient Survey \citep[CRTS;][]{Drake:09:CRTS},  Skymapper \citep{Keller:07:SkyMapperOverview}, PanSTARRS \citep{Kaiser:10:PanSTARRSSurvey}, the Palomar Transient Factory \citep[PTF;][]{Law:09:PTFOverview}, the All-Sky Automated Survey for Supernovae \citep[ASAS-SN;][]{Shappee:14:ASASSN}, the Asteroid Terrestrial-impact Last Alert System \citep[ATLAS;][]{Tonry:18:ATLAS}, the Korea Microlensing Telescope Network \citep[KMTNet;][]{Kim:16:KMTNET}, the Dark Energy Survey \citep{DESCollaboration:05:DES,DES:16:MoreThanDE} and other surveys using the Dark Energy Camera \citep{Flaugher:15:DECam}, Hyper Suprime-Cam \citep{Miyazaki:18:HSC,Aihara:18:HSCSSP}, and the Evryscope \citep{Law:15:EvryscopeScience},
surveys exploited these new capabilities to improve a subset of depth, areal coverage, filter selection, and/or time-domain sampling.

Here we present a new sky survey, the Zwicky Transient Facility (ZTF)\footnote{\url{http://ztf.caltech.edu}}.  
ZTF's new CCD camera for the first time fills the focal plane of the Palomar 48-inch Schmidt, providing three orders of magnitude improvement in survey speed relative to the photographic surveys, by virtue of higher QE and substantial reduction in time between exposures.
If it could ignore daylight, ZTF could repeat the entire POSS survey in one day.

This paper provides a general overview of the ZTF observing and data systems, describes the on-sky performance and public surveys, and presents initial results in transient, variable, and solar system science.
Additional papers discuss specific ZTF aspects in greater detail:
\citet{tmp_Graham:18:ZTFScience} describe the scientific objectives of ZTF.  
\citet{tmp_Dekany:18:ZTFObservingSystem} provide an in-depth description of the design of the observing system.
\citet{tmp_Bellm:18:ZTFScheduler} discuss the ZTF surveys and scheduler.
\citet{tmp_Masci:18:ZTFDataSystem} detail the ZTF data system.
\citet{tmp_Patterson:18:ZTFAlertDistribution} present the alert distribution system employed by ZTF.
\citet{tmp_Mahabal:18:ZTFMachineLearning} discuss several applications of machine learning used by ZTF.
\citet{tmp_Tachibana:18:PS1StarGalaxy} presents a new star/galaxy classifier developed for ZTF from the PanSTARRS DR1 catalog \citep{Chambers:16:PS1,Flewelling:16:PS1db}.
\citet{tmp_Kasliwal:18:GROWTHMarshal} describe a web-based interface used by the ZTF collaboration to identify, track, and follow up transients of interest.

\section{Observing System} \label{sec:observing_system}

The capability of a survey camera to discover astrophysical transients can be quantified by its volumetric survey speed: the spatial volume within which it can detect an object of given  absolute magnitude, divided by the total time per exposure \citep{Bellm:16:Cadences}.  
This quantity combines limiting magnitude, field of view, and exposure and overhead times into a single metric capturing how quickly a survey can probe physical space for new events.

The ZTF concept assumed reuse of the Palomar 48-inch Samuel Oschin Schmidt Telescope.
The subsequent design of the ZTF observing system---the  camera, telescope, and associated subsystems---then attempted to maximize the volumetric survey speed of the system within a fixed cost envelope subject to this constraint.
This goal required maximizing the field of view of the camera while maintaining image quality, minimizing beam obstruction, and minimizing readout and slew overheads.
The final design achieves more than an order of magnitude improvement in survey speed relative to PTF\footnote{ZTF's median overhead time is about 10.2\,s compared to 42.0\,s for PTF, which had median $R$-band limiting magnitudes of 20.7\,mag in 60\,s exposures.  
For a fiducial object with $M_r=-19$\,mag, then, $\dot{V}_{-19} =  3.5\times10^4$\,Mpc$^3$\,s$^{-1}$ for ZTF as built, a factor of 14.9 larger than for PTF \citep{Bellm:16:Cadences}.}.

\citet{tmp_Dekany:18:ZTFObservingSystem} describes the as-built observing system in greater detail.

\begin{table*}
\begin{center}
\caption{Specifications of the ZTF Observing System \label{tab:specs}}
\begin{tabular}{ll}
\hline
\multicolumn{2}{c}{Telescope and Camera} \\
\hline
Telescope & Palomar 48\,inch (1.2\,m) Samuel Oschin Schmidt\\
Location & 33\degr\,21\arcmin\,26\farcs35\,N, 116\degr\,51\arcmin\,32\farcs04\,W, 1700\,m  \\
Camera field dimensions & $7.50^\circ$ N-S $\times$ $7.32^\circ$ E-W\\
Camera field of view & 55.0\,deg$^2$ \\
Light-sensitive area & 47.7\,deg$^2$ \\
Fill factor & 86.7\% \\
Filters & ZTF-$g$, ZTF-$r$, ZTF-$i$ \\
Filter exchange time & $\sim$110\,sec, including slew to stow\\
Image quality & $g=2.1^{\prime\prime},~r=2.0^{\prime\prime},~i=2.1^{\prime\prime}$ FWHM \\
Median Sensitivity (30\,sec, 5$\sigma$) & $m_g = 20.8,~m_r = 20.6,~m_i = 19.9$\\
& $m_g = 21.1,~m_r = 20.9,~m_i = 20.2$ (new moon)\\
\hline
\multicolumn{2}{c}{CCD Array} \\
\hline
Science CCDs & 16 6144$\times$6160 pixel e2v CCD231-C6 \\
Guide and Focus CCDs & 4 2k$\times$2k STA; delta doped by JPL\\
Pixels & 15\,$\mu$m\,pixel$^{-1}$\\
Plate scale & $1.01^{\prime\prime}$\,pixel$^{-1}$ \\
Chip gaps & 0.205$^\circ$ N-S, 0.140$^\circ$ E-W \\
CCD readout channels & 4 \\
Readout time & 8.2\,seconds \\
Read noise & 10.3 e- (median)\\
Gain & 5.8 e-/ADU \\
Linearity & 1.02\% +- 0.09\%   (correction factor variation)\\
Saturation & 350,000 e-\\
\hline
\end{tabular}
\end{center}
\end{table*}

\subsection{CCD Mosaic}

\begin{figure}
\includegraphics[width=\columnwidth]{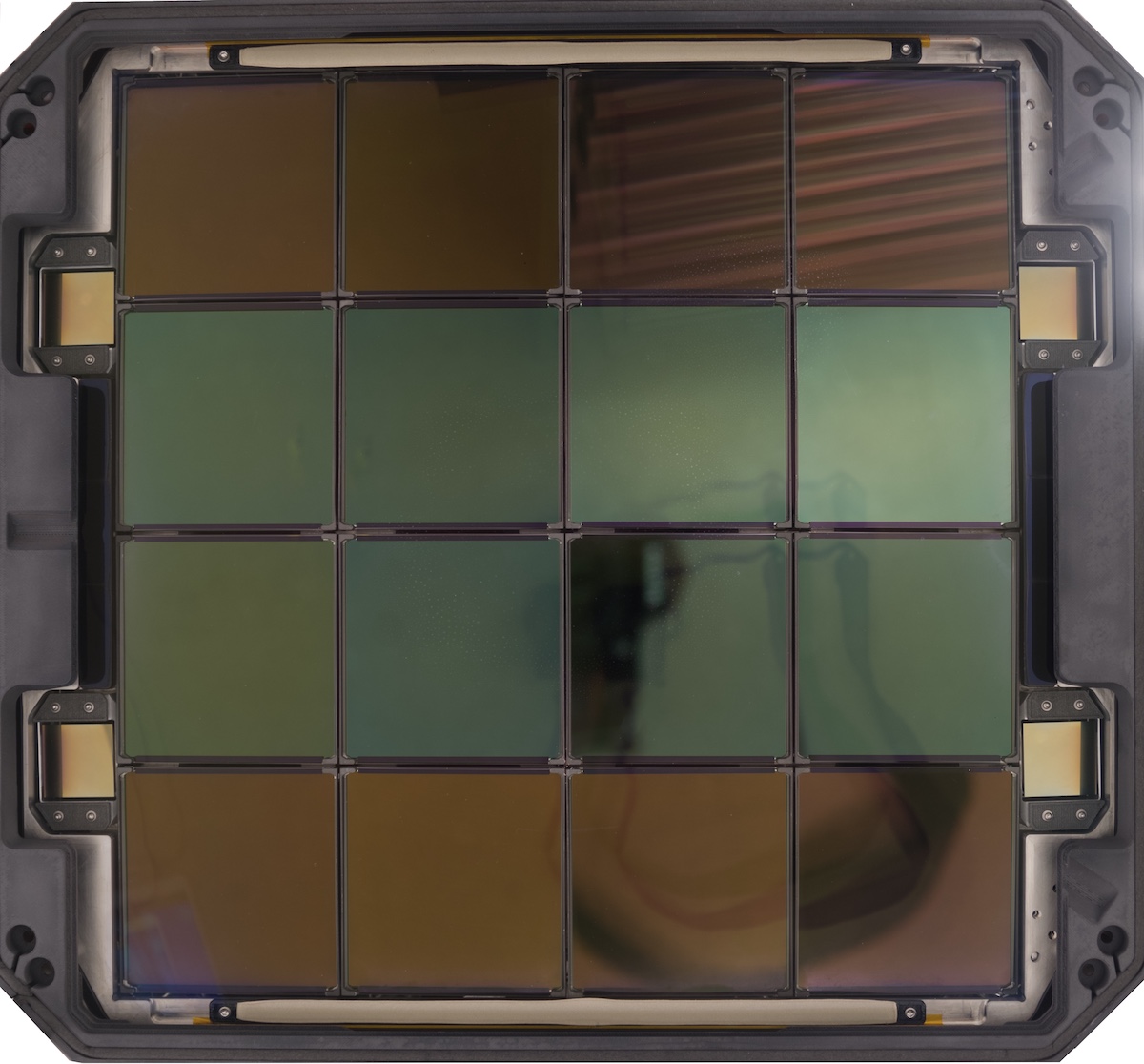}
\caption{Image of the ZTF focal plane.
The top and bottom rows of 6k $\times$ 6k science CCDs have a single-layer anti-reflective coating, while the middle rows have a dual-layer coating. 
Four 2k$\times$2k CCDs are located on the perimeter of the mosaic; one serves as a guider while the remaining three control tip, tilt, and focus.
North is up and East is left.
\label{fig:focal_plane}}
\end{figure}

\begin{figure}
\includegraphics[width=\columnwidth]{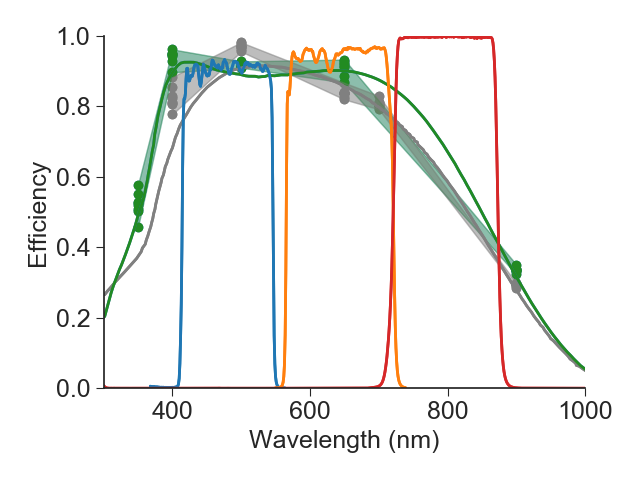}
\caption{On-axis filter transmission for the ZTF $g$, $r$, and $i$-band filters (blue, orange, and red lines).  
Grey and green points are measurements of the quantum efficiencies of the CCDs with single- and double-layer anti-reflective coatings, respectively.  Shaded regions show the range of these measurements, while grey and green lines show a model of the quantum efficiency for each configuration.
\label{fig:filters_qe}}
\end{figure}

The P48 was designed to use 14-inch square photographic plates, providing a field of view of 43.56\,deg$^2$ \citep{Harrington:52:P48}.  
Large-format ``wafer-scale'' CCDs proved the 
most cost-effective means of filling this large area and had the additional advantage of minimizing losses due to chip gaps. 
Our goal of maximizing throughput while minimizing cost motivated our decision to survey primarily in filters near the peak quantum efficiency of standard silicon.
We selected backside-illuminated standard silicon CCD231-C6 devices from e2v, Inc.  
The 15\,$\mu$m pixels provided critical sampling of the expected 2.0$^{\prime\prime}$ FWHM point spread function (\S \ref{sec:optics}) at a plate scale of 1.01$^{\prime\prime}$\,pixel$^{-1}$ while moderating data volume.  
(This pixel scale also matched that of the PTF camera.) 
Half of ZTF's CCDs have a single-layer anti-reflective coating, while the other half has a dual-layer coating that provides improved quantum efficiency in the $g$ and $r$ bands (Figures  \ref{fig:focal_plane} \& \ref{fig:filters_qe}).

The CCDs are nearly perfect cosmetically having only a few blocked columns. QE is uniform to a few percent on large scales.  Response non-uniformity on short scales is 0.55\% at 400 nm falling linearly to 0.3\% at 650nm.  Dark current is negligible in maximum exposure times contemplated (300 s). Well capacity is typically 350,000 e-, and charge transfer inefficiency is < 5ppm per pixel shift. 

Four 2k$\times$2k CCDs located around the perimeter of the mosaic serve as guide and focus sensors.  These  are STA-designed fully depleted thick CCDs that were delta-doped and multi-layer anti-reflection coated  by the JPL Micro Devices Laboratory.
Three are offset 1.45\,mm beyond the plane of science CCDs to allow determination of tip, tilt, and focus by computing the square root of the 2nd moment of the out-of-focus images. 
The fourth in-focus CCD is used for guiding.

\subsection{Cryostat}

Because the focus of a Schmidt telescope is located within the telescope tube itself, maximizing throughput requires minimizing the beam obstruction caused by the ZTF camera and related components.  
We located the readout electronics (\S \ref{sec:readout}), shutter (\S \ref{sec:shutter}), and filter exchanger (\S \ref{sec:filters}) outside the telescope tube.

\begin{figure}
\includegraphics[width=\linewidth]{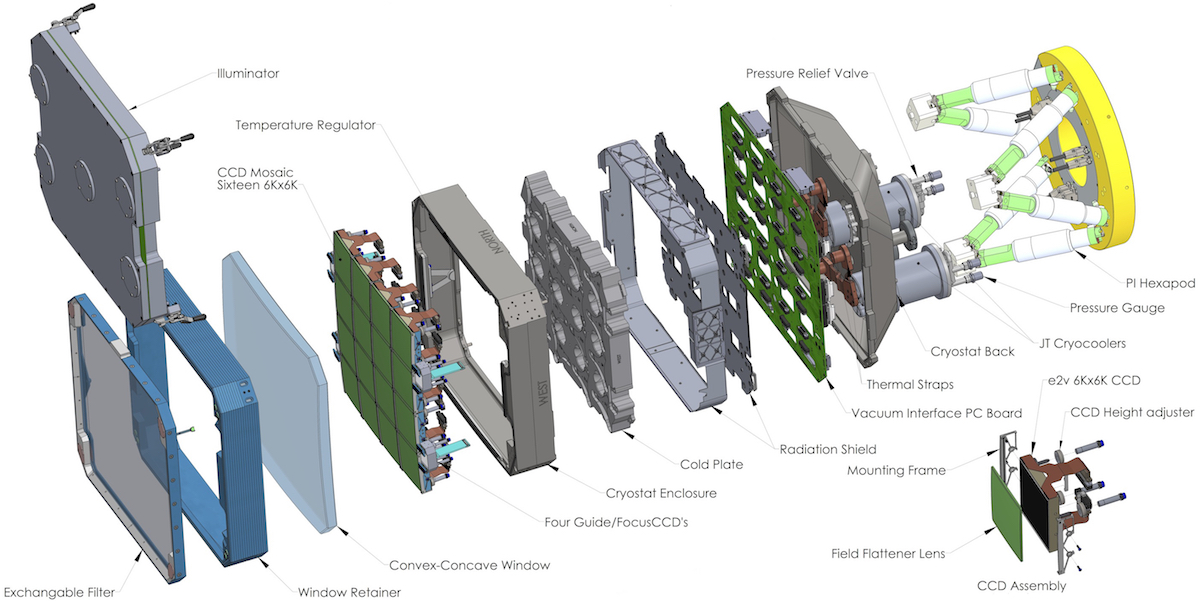}
\caption{Cutaway view of the ZTF cryostat.
\label{fig:cryostat_exploded}}
\end{figure}

\begin{figure}
\includegraphics[width=\linewidth]{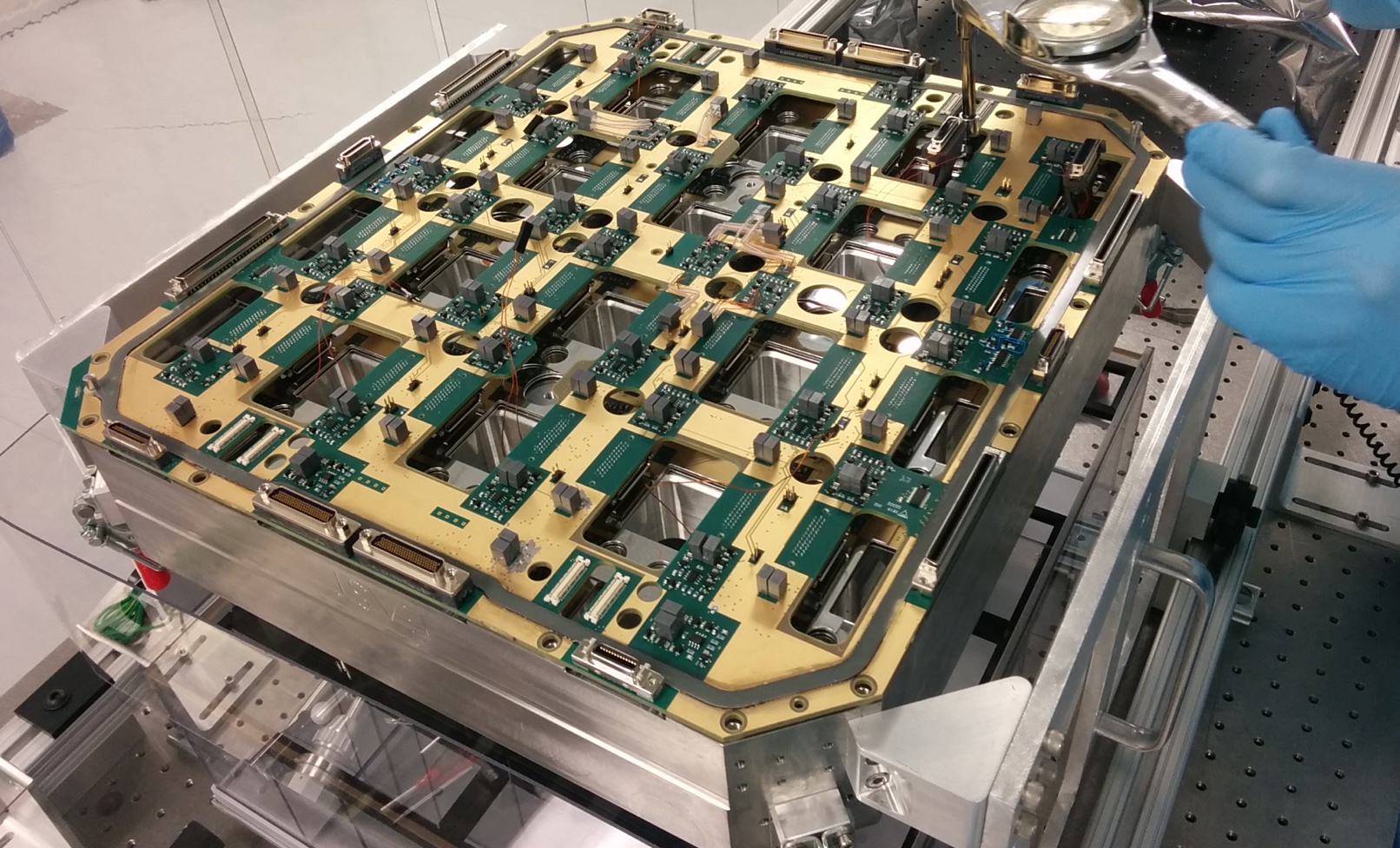}
\caption{Rear view of the vacuum interface board during cryostat assembly.  The vacuum gasket and connectors are visible around the perimeter.  
Holes in the interior provide space for the CCD flex cables as well as the control rods used during assembly. 
\label{fig:vib}}
\end{figure}

The cryostat can achieve its extraordinary compactness by a signal routing strategy based on a vacuum interface board, a printed circuit board having embedded traces and sandwiched between two O-rings that maintain vacuum (Figures \ref{fig:cryostat_exploded} \& \ref{fig:vib}).  On the outer edge of the interface board, commercial off-the-shelf connectors mount flush against the side of the cryostat, behind the beam footprint of the front window. Low obscuration (22.4\% including spiders) is achieved at the Schmidt prime focus despite the mosaic having comparable diameter to other major CCD cameras: 
The ZTF mosaic measures 560\,mm from corner to corner, similar in size to the Dark Energy Camera (525\,mm diameter) and about half of the area of the Large Synoptic~Survey Telescope camera (640\,mm diameter).

\subsection{Readout Electronics} \label{sec:readout}
  
Each four-CCD quadrant of the mosaic is operated completely independently by a sixteen-channel CCD controller, the ``Archon'' made by Semiconductor Technology Associates (STA) with 100 MHz video processor employing digital correlated double sampling.  A fifth Archon reads the three focus CCDs and guider though two channels each.  All controllers share a 100 MHz master clock and trigger to support the pixel-synchronous readout required to eliminate patterns that would be caused by  crosstalk from clocks on one controller to the video signal of another. The guide/focus CCDs cannot operate at the same speed, so one pixel is read for every three science pixels, to retain synchronization.  
True differential outputs of the science CCDs provide clock feed-through attenuation and crosstalk suppression, which in combination with clock slew rate minimization allows pixel time to be reduced to 830 ns \citep{tmp_Smith:18:SpeedNoise}.
Novel concurrent parallel clocking conceals line shift overhead so that readout time is only 8.2\,sec, while delivering 10.3 electrons median read noise, well below the minimum shot noise in the sky (27 electrons).

\subsection{Optics} \label{sec:optics}

Maintaining PTF's moderate image quality (2.0$^{\prime\prime}$ FWHM in $r$-band) over the larger ZTF focal plane required novel optics.  
The focal surface of the Schmidt telescope is curved; the glass planes used in the photographic surveys conducted with P48 were physically bent on a mandrel to conform to this shape \citep{Harrington:52:P48}.
For PTF, an optically powered dewar window was sufficient to provide good image quality over the flat CCD array. However, this approach alone was insufficient to correct the much larger field of view of ZTF.

The final ZTF optical design consists of four optically-powered elements as well as the flat filters (\S \ref{sec:filters}).  
In front of the existing achromatic doublet Schmidt corrector that was installed for the Second Palomar Sky Surveys \citep{Reid:91:POSS-II}, we installed a ``trim plate'' to modify the aspheric coefficient of the Schmidt corrector at the telescope pupil by about 10\%. The trim plate was figured by the Nanjing Institute for Astronomical and Optical Technologies (NIAOT) in China, from a Corning fused silica blank.
To handle the extreme field curvature of the Schmidt focus, the dewar vacuum window is a meniscus (with a conductive ITO coating on the inner surface providing resistive heating.) The CCDs themselves are mounted to a faceted cold plate, where each facet is a chord of the residual field curvature. Finally, to compensate for remaining curvature over each large science CCD, we mount $\sim$2 mm thick fused silica field flattener lenses 2 mm above each detector (Figure \ref{fig:cryostat_exploded}).

Ultimately, the useful field of the view of ZTF is limited by the Schmidt telescope design. At extreme field angles part of the beam from falls beyond the edge of the telescope primary mirror, with vignetting reaching 30\% in the corners. 

\subsection{Shutter} \label{sec:shutter}

To minimize beam obscuration within the telescope tube, we placed the shutter at the entrance pupil of the telescope.  This shutter was developed through a collaboration of Deutsches Elektronen-Synchrotron (DESY) and Bonn Shutter, who successfully delivered a bi-parting shutter with 1.2-meter aperture that opens and closes within 430\,msec while imparting less than 3 grams unbalanced force to the top of the telescope with negligible impact on image stability.

\subsection{Telescope}

In order to optimize survey productivity as a system, and increase reliability, we  invested in numerous upgrades to the Samuel Oschin Telescope.  To minimize slew overheads, we upgraded both telescope drive axes as well as the dome drive system to enable faster slews. After upgrades, the (hour angle, declination, dome) drive accelerates to and decelerates from a top speed of (2.5,~3.0,~3.0)$^\circ$\,s$^{-1}$ at 
(0.4,~0.5,~0.5)$^\circ$\,s$^{-2}$.  
With these upgrades, the telescope can slew and settle between adjacent fields, separated by 7 degrees, entirely during the CCD readout time. 

Other upgrades \citep{tmp_Dekany:18:ZTFObservingSystem} included a new three-vane instrument spider (to reduce beam obscuration), improved baffling of the telescope tube (to reduce scattered light), facility electrical improvements and lightning protection, a new dry air system (to inhibit condensation on window and refrigerant lines), refurbishment of the wind screen, and better thermal management in the dome.

\subsection{Filters and Filter Exchanger} \label{sec:filters}

ZTF has a complement of three custom filters, ZTF-$g$, ZTF-$r$, and ZTF-$i$. 
Given the differences of the ZTF system relative to potential calibrators (SDSS, PS1, \textit{Gaia}), we did not attempt to match any existing filter bandpasses exactly.  
Instead, we sought to maximize the signal-to-noise ratio achieved by avoiding major lines in the Palomar sky background and to control costs on the large filters.
Materion Precision Optics manufactured the $g$ and $r$ band filters and Asahi Spectra produced the $i$-band filter.


Our desire to minimize beam obstruction motivated an unusual design for the filter exchanger.  
We store all three filters in slots in a filter cabinet mounted in an access hatch of the telescope tube. 
A commercial robotic arm 
stows in a similar position. 
During the filter exchange, the arm uses a set of solenoid-deactivated magnets and redundant latches on its manipulator to dock with the frame holding the desired filter in the storage cabinet, move it to the camera, and secure it there.  
The arm then disconnects from the filter frame and stows against the wall of the telescope tube.
At present, for safety, the exchanges are only conducted when the telescope is in a quasi-horizontal stow position where none of the moving elements physically pass over the telescope primary mirror.
Including the slew time to and from the stow position, the additional overhead to change filters is $\sim$100\,seconds in typical operations. 
Additional optimization of the arm motion profile and exchanging closer to zenith is expected to reduce this further.

\subsection{Robotic Observing System}

The ZTF Robotic Observing Software (ROS) is based on the Robo-AO observing system \citep{Riddle:12,tmp_Dekany:18:ZTFObservingSystem} though many of the underlying tools were retained and upgraded.  ROS is based on a modular, fail-safe, multi-threaded, multi-daemon software architecture.  It has been designed to be able to run continuously for an extended period, while allowing human operators  to monitor the system to determine its  performance, track nightly errors, and reconfigure parameters if necessary.  Configuration files support engineering and science operation modes. Extensive telemetry is aggregated from all ZTF hardware and telescope control subsystems.

ROS is hosted on a single supervisory computer (which also controls the guide and focus CCDs) and four Archon camera control computers, each of which is responsible for readout of one quadrant of four science CCDs.  
Sufficient on-site data storage exists for at least two weeks of regular observing, in the unlikely event the microwave link from Palomar (\S \ref{sec:transfer}) were to suffer an outage.

\subsection{Scheduler}

The ZTF scheduler determines which fields to observe and in what order.
Integer Linear Programming techniques inspired by  \citet{Lampoudi:15:LCOGTScheduler} maximize the volumetric survey speed
using slot-based lookahead throughout the night.
\citet{tmp_Bellm:18:ZTFScheduler} describes the scheduling system in detail.

Due to the desire to simplify the data processing for image subtraction (\S \ref{sec:image_differencing}), all ZTF images are obtained on a fixed grid of fields with minimal dithering.  
The primary grid covers the entire sky with an average overlap between fields of about 0.26$^{\circ}$ in Declination and about 0.29$^{\circ}$ in Right Ascension.
The fields are aligned to cover the Galactic Plane region with the fewest pointings, improving the efficiency of both Galactic and extragalactic surveys.  
We also took care to ensure that large local galaxies were placed effectively.
A secondary grid of pointings, offset diagonally by about 5$^\circ$, fills in most of the CCD gaps and improves the fill factor within the survey footprint from 87.5\% to 99.2\% assuming no dithering.

\subsection{Flat Field Illuminator}

PTF constructed its flat fields from science images taken each night.  
In addition to preventing final reduction of the images until the end of the night \citep{Laher:14:PTFPipeline}, this scheme was negatively affected by fringing of sky lines and scattered light from the moon and other bright sources and proved to be among the factors limiting PTF's photometric precision.
For ZTF, a new Flat Field Illuminator system enables stable calibration frames to be taken before the night's observing.

The Flat Field Illuminator consists of a reflective screen, LED illuminators, and a baffled enclosure.
It is mounted on a tower close to the P48 dome.
Twenty four narrow-spectrum LEDs in each of 15 wavelengths spanning the ZTF filter bandpasses are spaced around a ring pointing towards a screen.
The screen is constructed from aluminum honeycomb paneling which makes it lightweight, stiff, and flat. Many coats of Avian-D white paint provide a very uniform lambertian scattering surface.  
The forward baffle mounted on the telescope docks to a similar baffle surrounding the flat field system to fully enclose the path between flat field screen and detectors.  
The enclosure walls are heavily baffled and covered with 2\% reflective Avian-D black paint facing the screen and black flocking 
facing the telescope.  
Similar baffles have been installed along the entire length of the enclosed telescope tube at sufficiently close spacing to block all single-bounce paths between flat field screen and primary mirror.  

This screen provides smooth and stable illumination for removing mid- to small-scale spatial frequencies in the sensitivity pattern. 
The 7\% radial intensity variation at the screen integrates to <2\% flat fielding error at the focal plane.  
This residual error occurs on large spatial scales that are easily corrected by calibrations derived from observing the relative photometry of stars as they are moved across the field (``star-flats'').

LEDs are driven by constant current sources, and their forward voltage is monitored to sense junction temperature so that temperature compensation can be applied if required. 
Flats are acquired separately in each LED wavelength and then combined with a relative weighting which minimizes the manifestation of CCD QE patterns in the "star-flats" which should only show mosaic-scale patterns.

The principal error observed in flats is a 6\% increase close to East and West edges of the CCDs where light scatters off the frames holding the field flattener lenses.  This additive background must be removed from flats since it does not represent enhanced sensitivity.  Fortunately it rises rapidly close to the edge of the CCD and can be fitted with sub-percent accuracy.

\section{Data System} \label{sec:data_system}

The ZTF data processing system is housed at the Infrared Processing and Analysis Center (IPAC) and builds on the lessons learned processing data from PTF and iPTF \citep{Ofek:12:PTFPhotometricCal,Laher:14:PTFPipeline,Masci:17:PTFIDE}.  
\citet{tmp_Masci:18:ZTFDataSystem}
provides a complete description of the ZTF pipelines.

\subsection{Data Transfer} \label{sec:transfer}

The CCD controllers sample the video signal at  100 MHz and 16 bit resolution, averaging multiple samples to produce a floating point output with about 18 bits of dynamic range. 
We use the \texttt{fpack} program \citep{Pence:10:Compression} to compress each quadrant and each overscan separately, allowing the compression to be optimized for the measured width of the core of the histogram in each image.  
In practice this noise root variance $\sigma$ is dominated by sky noise ($\geq 25$\,e$^{-}$\,s$^{-1}$\,pixel$^{-1}$).  
\texttt{fpack} converts the floating point data to integers applying a normalization factor $q = 2$, which results in $\sigma = 2$ for the histogram of integers.  
Lossless Rice compression is then applied.  
We apply a pseudo-random dither prior to normalization to avoid biases produced by rounding.  
The same dither values are subtracted after decompression (using the \texttt{funpack} program) so that the dither does not add noise.  
The result is that number of bits per pixel is reduced to an average of 5 during data transport at a cost of a 1\% increase in sky noise, due to quantization by the normalization step. 
Despite the slight increase in noise, our tests confirm that this procedure does not appreciably bias image coaddition or photometry \citep[cf.][and references therein]{PriceWhelan:10:ImageBandwidth,Pence:10:Compression}.

The observing system transfers the images to IPAC via the NSF-funded High Performance Wireless Research and Education Network (HPWREN) administered by the University of California San Diego.  Typical transfer times are <25\,seconds, sufficient to keep up with the fastest  observing cadence (38.3\,seconds) throughout the night.

\subsection{Image Processing} \label{sec:image_processing}

Upon arrival, each multi-extension FITS image\footnote{\url{https://fits.gsfc.nasa.gov/}} is split into four readout quadrants per CCD and farmed out in parallel to the processing cluster.
All subsequent processing is conducted independently on each CCD readout quadrant.
Each image is tagged with the observing program that obtained it (public, collaboration, or Caltech), and the access permissions for all of the downstream data products are propagated accordingly.

The image processing pipeline first subtracts bias frames and applies the flat field correction.  
The pipeline then calls the \texttt{SCAMP} package \citep{Bertin:06} to determine a World Coordinate System using \textit{Gaia} DR1 \citep{Brown:16:Gaia1} as the reference catalog.  
Subsequently the pipeline fits a zero point and color term to  photometrically calibrate the quadrant to PanSTARRS 1 \citep{Chambers:16:PS1}.
The pipeline sets appropriate mask bits for saturation, bad pixels, ghosts, and other instrumental artifacts.

The pipeline produces both point-spread function (PSF) fit  (DAOPHOT, \citealp{Stetson:87}) and aperture (SExtractor, \citealp{Bertin:96}) photometry catalogs from the processed direct image, and the raw and processed images and catalogs are archived (\S \ref{sec:archive}).

\subsection{Reference Image Generation}

Coadded reference images are required for image differencing (\S \ref{sec:image_differencing}) as well as lightcurve source association (\S \ref{sec:lightcurves}).
We construct reference images for each field, filter, and quadrant combination. 
Typical stacks have at least 15 images.
We use \textit{Swarp} \citep{Bertin:02} to map the images to a common footprint and then compute an outlier-rejected average.
Reference building pipelines are triggered automatically at the end of the night. 

\subsection{Image Differencing} \label{sec:image_differencing}

The image differencing pipeline identifies moving and changing sources.  
It first prepares the processed science image and reference image by matching their photometric throughputs, warps the reference image onto the science image, and matches their backgrounds at low spatial frequencies. 
PSF-matching, image differencing, and the creation of an accompanying match-filtered image optimized for detecting point sources on the difference are then performed using the ZOGY algorithm \citep{Zackay:16:ZOGY}. 
The pipeline then detects both positive and negative ``candidate'' sources at a signal-to-noise ratio greater than 5.
The pipeline also measures a variety of pixel-based features for each candidate (e.g., the number of positive and negative pixels in a region around the candidate) to provide to the Real-Bogus machine learning algorithm \citep{tmp_Mahabal:18:ZTFMachineLearning}.
Each candidate is loaded into a database and then packaged with other contextual information into an alert packet (\S \ref{sec:alert_stream}) for distribution.

The realtime pipeline runs from raw images to transient alerts in about four minutes.

\subsection{Transient Alert Stream} \label{sec:alert_stream}

The ZTF alert distribution system provides near-real-time access to transient and variable events identified by the image differencing pipelines.
To aid the user in filtering the full alert stream for sources of interest, the ZTF alert stream provides rich alert packets containing not only the measurement that triggered the alert, but also a wide variety of contextual information.  
These include  a Real-Bogus score \citep{tmp_Mahabal:18:ZTFMachineLearning} assessing the probability the candidate is astrophysical, a
lightcurve of previous detections (or upper limits) from the last 30 days, a summary of prior detections before the 30-day window, cross-matches to the Pan-STARRS1 catalog along with a probabilistic star-galaxy score \citep{tmp_Tachibana:18:PS1StarGalaxy}, and FITS cutouts of the science, reference, and difference images. 

The alert packets themselves are serialized in the open-source Apache Avro format\footnote{\url{https://avro.apache.org}}.
Schemas, example packets, and complete documentation of the packet fields are available\footnote{\url{https://github.com/ZwickyTransientFacility/ztf-avro-alert}}.

The alert packets are distributed using the open-source queue system Apache Kafka\footnote{\url{https://kafka.apache.org/}}.
Kafka provides a distributed queue that is scalable to the alert volumes expected by LSST.
\citet{tmp_Patterson:18:ZTFAlertDistribution}
describes the architecture and implementation of the alert distribution system more fully. 

Alerts from ZTF's public survey stream in near-real time to community brokers, including the Arizona-NOAO Temporal Analysis and Response to Events System \citep[ANTARES;][]{Narayan:18:ANTARES}, ALeRCE\footnote{Automatic Learning for the Rapid Classification of Events; \url{http://alerce.science/}}, Lasair, and Las Cumbres Observatory\footnote{\url{https://mars.lco.global}} which will provide public access.
While the community brokers come online, we are also providing a bulk nightly release of public alerts\footnote{\url{https://ztf.uw.edu/alerts/public/}}.

\subsection{Solar System Processing}

Solar System Processing is divided between searches for streaked Near-Earth Objects and point-like moving objects.
Both are detected in the difference image processing.  
Streaked objects are identified by a dedicated pipeline
originally developed for PTF \citep{Waszczak2017}. 

Point-like moving object candidates are identified at the end of the night by the ZTF Moving-Object Discovery Engine (ZMODE).  
ZMODE attempts to link tracklets from the last three observing nights and then fit orbits to them.
High-quality objects are forwarded to human scanners for vetting and then reported to the Minor Planet Center.

\subsection{Direct Imaging Lightcurves} \label{sec:lightcurves}

For archival studies of variable stars and AGN in uncrowded fields, lightcurves built from direct (un-subtracted) images provide a higher-fidelity data product because they avoid the subtraction artifacts and additional noise produced by difference imaging.
We build lightcurves every few months from the calibrated PSF photometry catalogs produced from the unsubtracted epochal images (\S \ref{sec:image_processing}).
Starting from the catalogs built from the deep reference images, we associate the sources in each epochal PSF photometry catalog with the nearest source in the reference catalog.  
The resulting lightcurves are stored in HDF5 ``matchfiles'' on a field, quadrant basis\footnote{This choice eases processing but means that photometry from the same source can appear in multiple files if observations are taken in the secondary pointing grid or if a source is near the readout quadrant boundary.}.
To further improve the photometric precision, we solve for a small per-epoch shift in the absolute calibration zeropoint by minimizing the scatter of non-varying stars \citep{Ofek:11:RelativePhotometry}, achieving better than 10\,mmag repeatability for bright, unsaturated sources.
Additionally, we store a variety of timeseries features \citep[cf.][]{Richards:11:MLClassifier} computed on the lightcurve to aid in identification of variable sources.

\subsection{Archive and Data Releases} \label{sec:archive}

The Infrared Science Archive (IRSA) at IPAC provides archival access to ZTF images, catalogs, lightcurves, and archived alert packets\footnote{See \url{http://irsa.ipac.caltech.edu/Missions/ztf.html}.  
PTF and iPTF data are publicly available through a comparable interface, \url{http://irsa.ipac.caltech.edu/Missions/ptf.html}}.
Both interactive web-based and programmatic queries are supported.
The first release of data products (other than alerts) from the public surveys is planned for one year after the start of the survey, in the second quarter of 2019\footnote{Note that members of the ZTF collaboration are not allowed to access archived data from the public surveys prior to the data release.}.

\section{On-Sky Performance} \label{sec:performance}


ZTF achieved first light in October 2017.  
Commissioning activities continued through March 2018 and combined technical activities to verify the performance of the observing and data systems with science validation experiments.

Formal survey operations began on March 20, 2018, although routine operations of the filter exchanger and guide and focus CCDs occurred only in April and June 2018 respectively.

\begin{figure}
\includegraphics[width=\columnwidth]{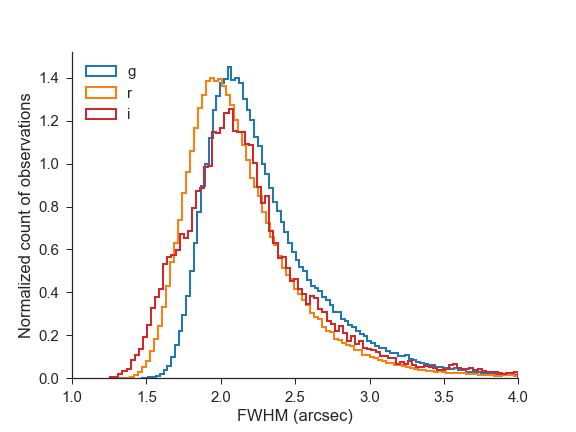}
\caption{Normalized histogram of point-source full-width at half-maximum (FWHM) for all images in $g$ (blue), $r$ (orange), and $i$ (red) bands during June 2018.
\label{fig:fwhm}}
\end{figure}

Figure \ref{fig:fwhm} shows the delivered image quality for all three ZTF filters.
Median image quality for the subset of observations above airmass 1.2 was 
2.1$^{\prime\prime}$ FWHM in $g$-band,
2.0$^{\prime\prime}$ FWHM in $r$-band, and 
2.1$^{\prime\prime}$ FWHM in $i$-band.

\begin{figure*}
\includegraphics[width=\textwidth]{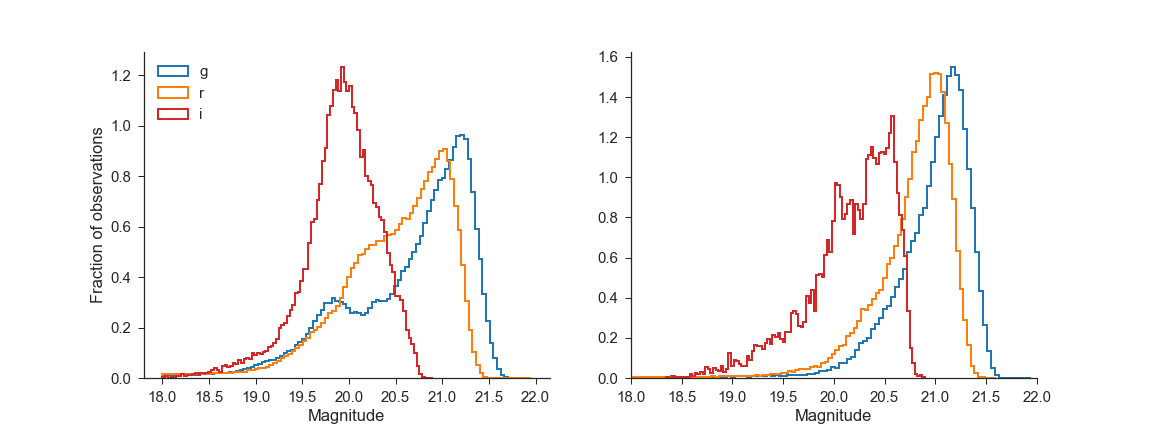}
\caption{Left: Histogram of five-sigma limiting magnitudes in 30 second exposures for $g$ (blue), $r$ (orange), and $i$ (red) bands over one lunation.
Right: Limiting magnitudes for observations obtained within $\pm$3\,days of new moon.
\label{fig:limiting_mag_hist}}
\end{figure*}

Figure \ref{fig:limiting_mag_hist} shows the limiting magnitudes obtained in all three filters over one lunation.  
Median five-sigma model limiting magnitudes are 
20.8\,mag in $g$-band, 20.6\,mag in $r$-band, and 19.9 in $i$-band.
Restricting to $\pm$3\,days around new moon,  the dark-time median limiting magnitudes are
21.1\,mag in $g$-band, 20.9\,mag in $r$-band, and 20.2 in $i$-band.

\section{Survey Strategy} \label{sec:survey_strategy}

ZTF divides its observing time between three high-level programs: public surveys (40\%), ZTF collaboration surveys (40\%), and Caltech surveys (20\%).
Each program in turn divides its time between multiple sub-surveys.
All of the available surveys are interleaved simultaneously by the survey scheduler \citep{tmp_Bellm:18:ZTFScheduler}, which optimizes each night's schedule for volumetric survey speed while maintaining balance among the programs.
Private surveys are not allowed to use the observation history of the public surveys in the scheduling process.
In addition to performing the regularly scheduled surveys, the scheduler can perform Target of Opportunity (TOO) observations in response to external triggers.
Each image is taken for one and only one owner in order to simplify 
access to derived data products (images, catalogs, lightcurve points, alerts).  
As the public surveys cover the entire available sky, some ``duplicate'' observations are unavoidable.
Here we give an overview of the public surveys; a detailed discussion of the surveys and on-sky scheduler performance will appear in \citet{tmp_Bellm:18:ZTFScheduler}.
\citet{tmp_Graham:18:ZTFScience} provides an overview of some of the expected scientific returns.

During its public time, ZTF is conducting the two surveys of broad scientific utility that we proposed to the NSF Mid-Scale Innovations Program (MSIP): a Northern Sky Survey and a Galactic Plane Survey.
Motivated by the LSST baseline cadence \citep[e.g.,][]{ivezic2008lsst}, the Northern Sky Survey is a three-day cadence survey of all fields with centers north of $\delta = -31^\circ$, except those in the Galactic Plane Survey\footnote{As of this writing, limits of the Telescope Control System exclude observations north of $\delta = 80^\circ$.}.  
The Galactic Plane Survey is a nightly survey of all visible fields in the region $|b| < 7^\circ$, $\delta > -31^\circ$.
For both surveys, each night a field is observed, it is visited twice, once in $g$-band and once in $r$-band, with at least 30\,minutes separation between the two visits \citep[cf.][]{Miller:17:ColorMeIntrigued}.
We expect to run these public surveys for at least the first eighteen months of the ZTF survey.

We will attempt to obtain low-resolution spectra for all likely extragalactic transients brighter than 18.5\,mag using the SED Machine \citep{Blagorodnova:18:SEDM} on the Palomar 60-inch and will publicly report these classifications \citep{Fremling:18:BTS}.

\section{First Results} \label{sec:first_results}

ZTF will enable new discoveries of many classes of astrophysical objects, including explosive extragalactic transients, optical counterparts of multi-wavelength and multi-messenger phenomena, variable stars, Tidal Disruption Events, Active Galactic Nuclei, 
and solar system objects.
\citet{tmp_Graham:18:ZTFScience} presents ZTF's science goals in detail.
In this section we present initial results in these areas from the early ZTF survey.

\subsection{Transient Science} 

During commissioning of the alerts system, we searched the incoming alerts for astrophysical transients, both providing feedback for the machine learning by marking ``bogus'' sources, and flagging potential supernovae for follow-up. 
Transient alerts were filtered and vetted via the GROWTH marshal system \citep{tmp_Kasliwal:18:GROWTHMarshal} and using a machine-learning based
classifier \citep{tmp_Mahabal:18:ZTFMachineLearning}.
In two months of commissioning data, we classified a total of 38 supernovae. 
Of these, 15 were only discovered by ZTF, while another 13 were first discovered by ZTF and later picked up by other surveys. 
The relatively modest yield is expected due to the limited set of reference images available, the need to maintain high thresholds while training the Real/Bogus system, and poor winter weather.
All classified supernovae from commissioning data have been made public on the Transient Name Server (see \citealp{Kulkarni:18:ATel11266} and \citealp{Lunnan:18:ATel11567} for details). The classification spectra are available on WISeREP \citep{yg12}.

\begin{figure*}
\begin{center}
\begin{tabular}{cc}
\includegraphics[width=3.5in]{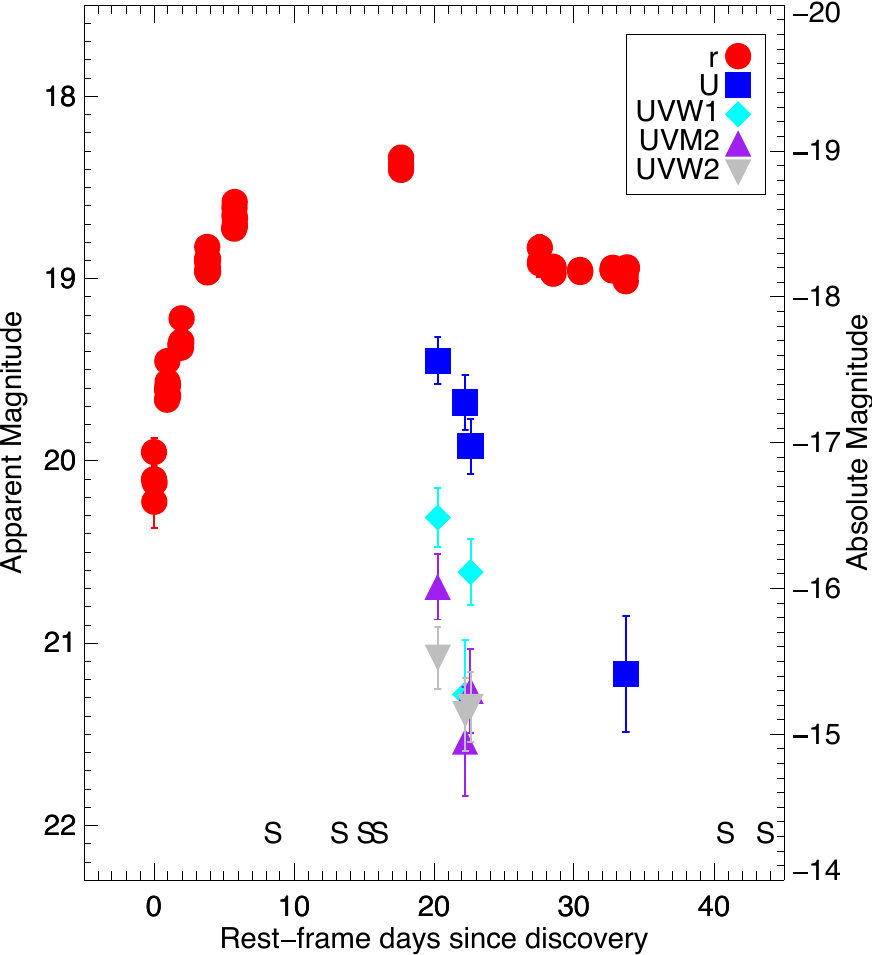} & \includegraphics[width=3.5in]{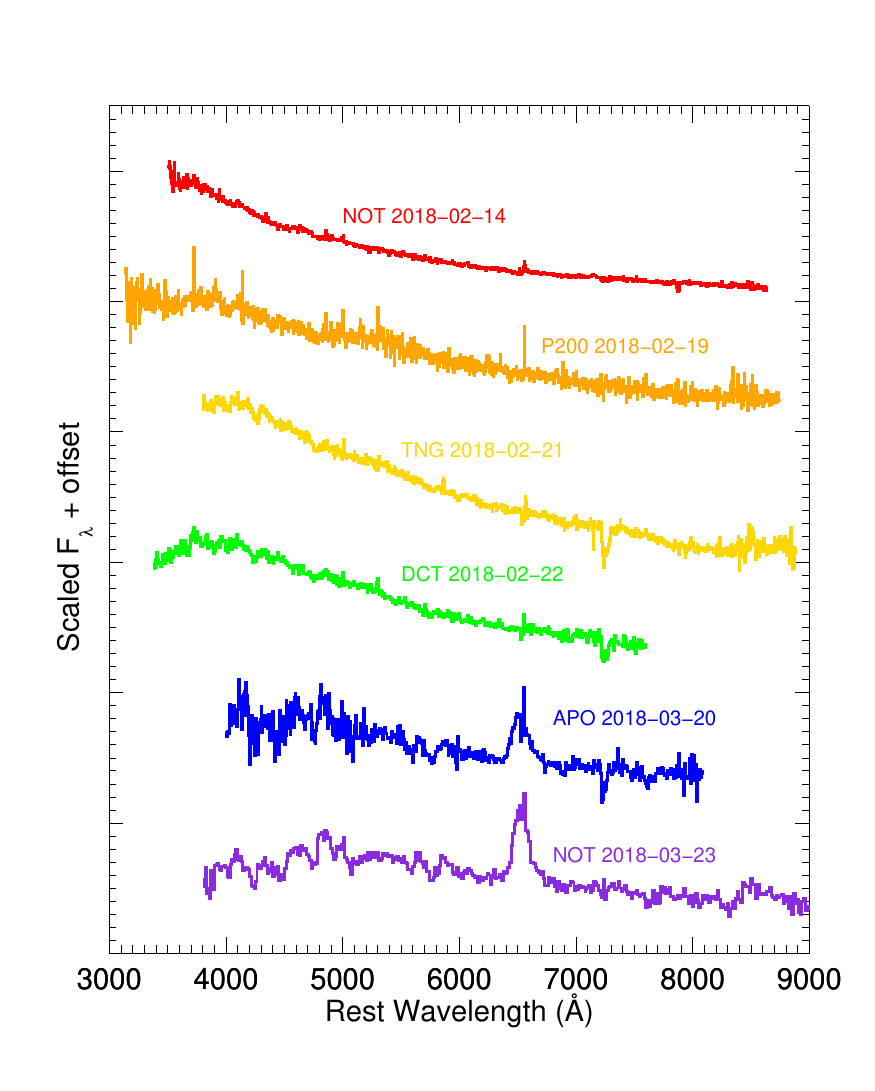}
\end{tabular}
\caption{Left: Light curve of ZTF18aaayemw. The rise is well captured in ZTF data. 20 days after discovery, the supernova is still detectable in the UV. The spectroscopic epochs are marked along the bottom axis. Right: Sequence of spectra of ZTF18aaayemw. The spectrum stays featureless and blue for at least the first $\sim 20$~days, before finally developing broad H$\alpha$ classifying ZTF18aaayemw as a SN~II.
\label{fig:snspec}}
\end{center}
\end{figure*}

As an illustrative example from the commissioning data, we present ZTF18aaayemw (SN\,2018yt), one of the first objects found, and one not identified by any other surveys. 
ZTF18aaayemw was discovered as a rising transient on 2018-02-07.26 (UT). 
Because this object was discovered so early in the survey, flux is also seen in the reference image that was built from data taken over the previous nights, so we cannot constrain the explosion date exactly. 
The light curve is shown in the left panel of Figure~\ref{fig:snspec}. 
An initial spectrum taken with the Nordic Optical Telescope on 2018-02-14 shows a featureless, blue continuum indicating a blackbody temperature of $\sim12,000$~K; narrow emission lines from the host galaxy sets the redshift at $z=0.0512$. 
We continued to follow ZTF18aaayemw, and the sequence of spectra obtained is shown in the right panel of Figure~\ref{fig:snspec}. 
The spectrum remained blue and featureless for at least two weeks after discovery; spectra taken a month later show broad H$\alpha$ emission, classifying ZTF18aaayemw as a SN~II. 
Details of the data collection and reduction are found in Appendix~\ref{sec:SNdetails}.

The early spectral evolution of ZTF18aaayemw is similar to that of other SNe~II such as SN~IIb iPTF13ast \citep{galyam14} and SN~IIn iPTF11iqb \citep{smith15}, which also did not show broad features until later than 15~days post-explosion. 
These two supernovae also showed flash spectroscopy features (i.e., features from the stellar envelope or circumstellar material ionized by the supernova shock breakout), which we do not observe in ZTF18aaayemw. 
This could be because no such features were present, or because they have faded by our earliest spectrum at $>7$~days. 
In the compilation of \citet{khazov16}, only 1/13 SNe~II where the first spectrum was taken 7-9 days after explosion showed flash features, while 3/13 showed blue, featureless continua like we see in ZTF18aaayemw.

\subsection{Target of Opportunity and Multi-Messenger Science}

We performed Target-of-Opportunity (ToO) follow-up observations in response to IceCube-171106A \citep{IceCube:17:GCN22105}, a neutrino of likely astrophysical origin with an estimated energy in excess of 1 PeV. The neutrino was detected by the IceCube Neutrino Observatory, and was distributed as part of the IceCube Realtime Program \citep{Aartsen:17:IceCube}. It was well-localized, with a sub-degree angular resolution, and was followed-up by ZTF with single-pointing observations. 
With ZTF's large field-of-view, such events will typically be covered by observations in a single field. 
Though the field was observed in this commissioning phase multiple times over a period of days, comparisons to reference images did not reveal any optical counterpart. Nonetheless, this example illustrates the potential of the ZTF ToO program to undertake multi-messenger observations of neutrino and gravitational-wave events. 

ZTF also observed the localization region of the short gamma-ray burst GRB180523A (trigger 548793993) detected by \textit{Fermi}-GBM.  
ZTF obtained a series of $r$ and $g$-band images covering 2900 square degrees beginning at 3:51 UT on 2018 May 24 (9.1 hours after the burst trigger time), corresponding to approximately 70\% of the probability enclosed in the localization region. 
Images in $r$ and $g$ bands were again taken the following night. More than 100 high-significance transient and variable candidates were identified by our pipeline in this area, all of which had previous detections with ZTF in the days and weeks prior to the GRB trigger time.  No viable optical counterparts were thus identified. The median 5 sigma upper limit for an isolated point source in our images was $r > 20.3$\, and $g > 20.6$\,mag. 

\subsection{Variable Science} 

During commissioning we also validated ZTF's utility for studying variable stars using direct (non-difference) imaging.

\subsubsection{Variability of Be Stars}
\begin{figure}
\includegraphics[width=3.5in]{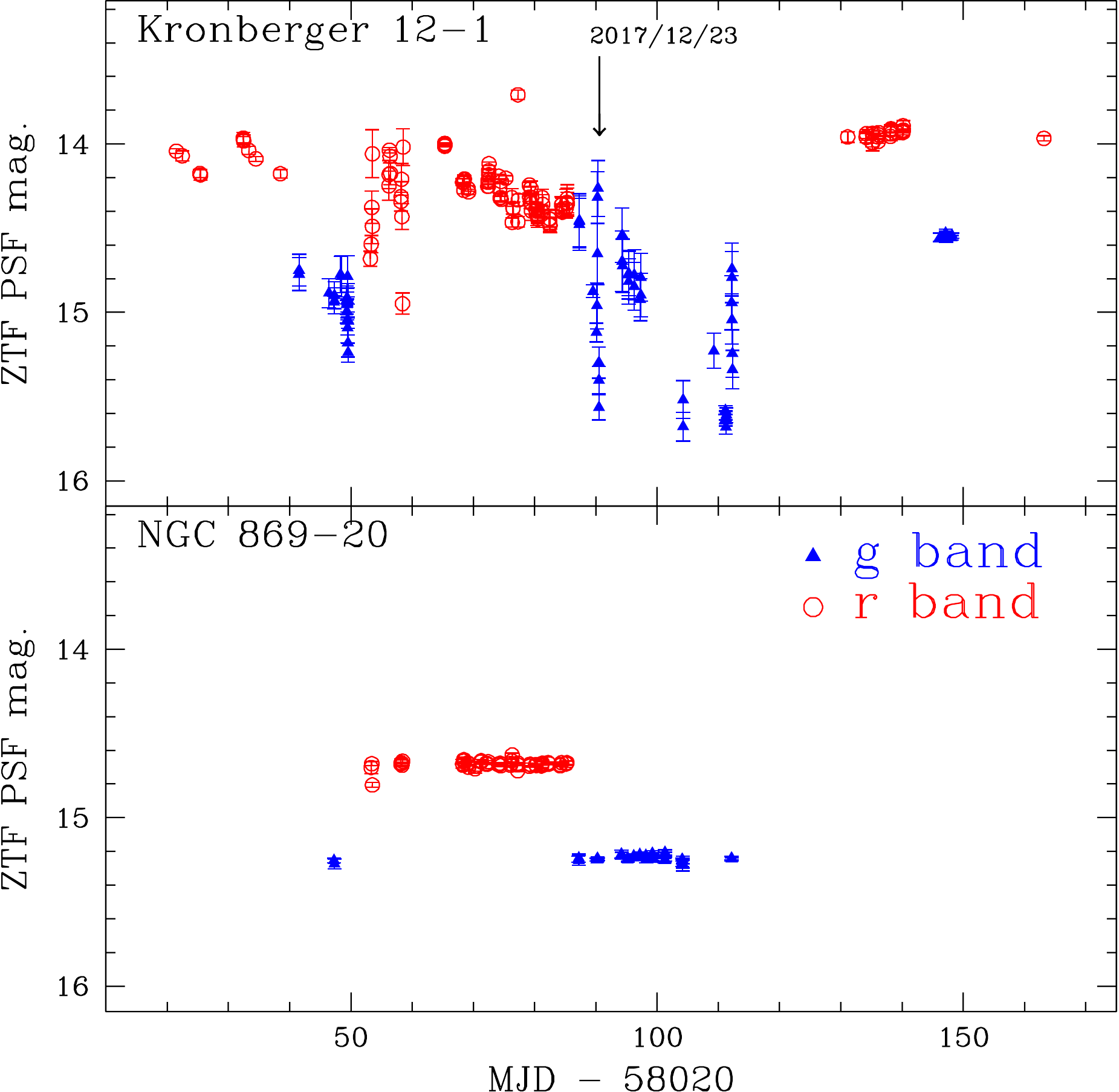}
\caption{ZTF light curves of two Be star candidates selected from \citet{Yu2018} in $g$ (filled blue triangles) and $r$ (open red circles) bands. The magnitudes are based on the PSF photometry but have not had relative photometry corrections applied \citep[see][]{tmp_Masci:18:ZTFDataSystem}, leading to larger observed scatter on a handful of nights. 
\label{fig:Be}}
\end{figure}

A fraction of Be stars are known to exhibit photometric variability due to the non-radial pulsation, ejected material, stellar winds, or instability of the decretion disk \citep[see review in][and references therein]{2013A&ARv..21...69R}. 
A variety of kinds of variability with different time scales have been reported, including outbursts, long-term variation, and periodic variations \citep{1997A&A...318..548O, 1998A&A...335..565H, Labadie2016}.
Using the ZTF commissioning data, we explored the variability on timescales of days to months of 83 Be star candidates in open clusters selected from \citet{Yu2018}. 
In our preliminary examination of these data, we found that less than $\sim10\%$ of our Be star candidates show qualitative variability. 
Figure~\ref{fig:Be} gives one example of a Be star candidate that exhibits variability (upper panel) and another one that does not (lower panel). 
We expect that a longer time baseline as well as further refinements of the lightcurve pipeline will provide valuable constraints on the variability of Be stars (such as variable fraction, amplitude of variation, outburst activity, and so on).

\subsubsection{RR Lyrae}

\begin{figure}
\includegraphics[width=3.5in]{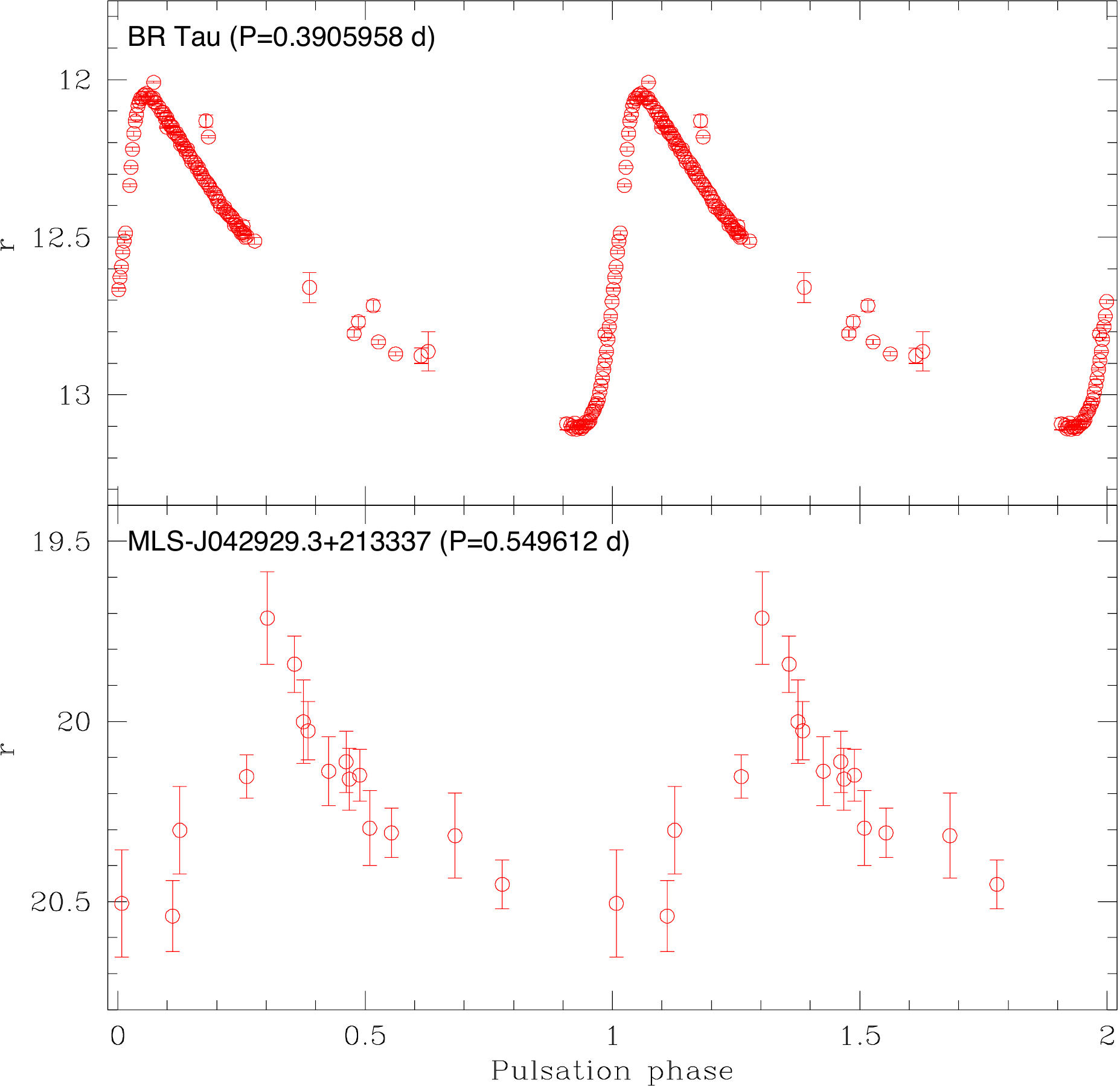}
\caption{ZTF $r$-band light curves for two known RR Lyrae based on PSF photometry but without relative photometric correction \citep[see][]{tmp_Masci:18:ZTFDataSystem}. The pulsation periods $P$ are taken from literature and not derived from the ZTF light curves.
\label{fig:RRab}}
\end{figure}

The homogeneous $gri$-band light curves for RR Lyrae provided by the ZTF are also a useful tool to investigate their pulsational properties. 
For example, the period-color and amplitude-color relations of RR Lyrae can be used to probe the interaction of photosphere with the hydrogen ionization front in these type of pulsating stars \citep[e.g.,][and references therein]{2017ApJ...834..160N}.
To check the light curve quality for large-amplitude variable stars such as RR Lyrae, we constructed the light curves of known RR Lyrae in one ZTF field based on the ZTF commissioning data. 
Figure~\ref{fig:RRab} shows the saw-tooth shape light curves for one bright and one faint RR Lyrae located in the selected ZTF field, demonstrating the expected light curve quality when ZTF is in full science operation. 
The finding of faint ($\sim20.5$~mag), and hence distant, RR Lyrae will be useful for the study of the Galactic halo \citep[e.g.,][and references therein]{Cohen2017}.

\subsection{Small Bodies in the Solar System}

Small solar system bodies encompass comets and asteroids, Trojans, Centaurs, near-Earth objects (NEOs), and trans-Neptunian objects. 
ZTF will provide extensive observations of thousands of small bodies, allowing long-duration measurements of their positions, motions and brightnesses as a function of time.  
Below we discuss the potential science return from the ZTF observations of solar system objects, and highlight four examples of early results from the first months of operation.

\subsubsection{Near Earth Objects}

The NEO search activity of ZTF comprises two components: detection of point-like NEOs, and detection of natural fast-moving objects that are moving more than a few degrees per day and hence appear as streaks. 
The ZTF Data System \citep{tmp_Masci:18:ZTFDataSystem} scans all ZTF difference images for these two types of objects and releases candidate detections in near real-time. 
Screening of new detections and submission to the Minor Planet Center (MPC) has been done on a best effort basis since February 2018 for those fields for which good reference images exist. 
On a clear night with cadence and fields suitable for asteroid detection, ZTF can produce $\sim100,000$ detections of $\sim25,000$ asteroids.

By 2018 May 4, after three months of operation, ZTF had submitted $\sim600,000$ measurements to the MPC and been assigned designations for about 320 new objects. 
The new discoveries include seven Near-Earth Asteroids
(Table~\ref{tbl:ztf-nea}), of which one (2018 CL) is a Potentially Hazardous Asteroid---an object with a minimum orbit intersection distance with Earth of less than 0.05\,A.U. and $M_H < 22$.
Five of these seven new NEOs were detected by the dedicated streak-detection pipeline \citep{Waszczak2017}. 
Current efforts are aimed towards optimizing this pipeline for better rejection of false positives as we better characterize the new camera and detectors, 
and using citizen science through Zooniverse to increase the size of the training sample \citep[for details, see][]{tmp_Mahabal:18:ZTFMachineLearning}.
Efficient new algorithms are also under development \citep{tmp_Nir:18:StreakDetection}.

\begin{table*}
\begin{center}
\caption{Near-Earth Asteroids discovered by ZTF as of 2018 April 30.\label{tbl:ztf-nea}}
\begin{tabular}{ccccc}
\tableline\tableline
Designation & Date of discovery & Orbit type & Discovery engine & Reference \\
\tableline
2018 CL & 2018 Feb 5 & Aten & Streak & \citet{MPECC23, Ye18} \\
2018 CP$_2$ & 2018 Feb 9 & Apollo & Point-source & \citet{MPECC73} \\
2018 CZ$_2$ & 2018 Feb 9 & Apollo & Point-source & \citet{MPECC88} \\
2018 GN$_1$ & 2018 Apr 10 & Apollo & Streak & \citet{MPECG56} \\
2018 GE$_2$ & 2018 Apr 10 & Apollo & Streak & \citet{MPECG73} \\
2018 HL$_1$ & 2018 Apr 21 & Apollo & Streak & \citet{MPECH70} \\
2018 HX$_1$ & 2018 Apr 23 & Apollo & Streak & \citet{MPECH80} \\
\tableline
\end{tabular}
\end{center}
\end{table*}

\subsubsection{Asteroid Light Curves}
\begin{figure*}
\plotone{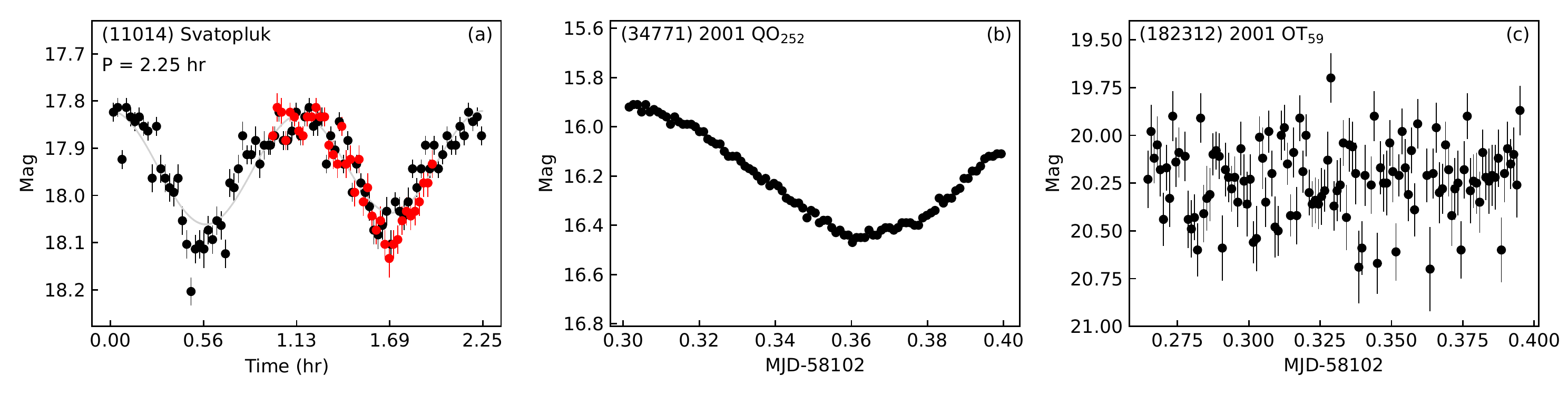}
\caption{ZTF $r$-band light curves of asteroid (11014) Svatopluk, (34771) 2001 QO$_{252}$, and (182312) 2001 OT$_{59}$.}
\label{asteroid}
\end{figure*}

Asteroid light curves obtained from high-cadence observations can secure the measurements of their rotation periods and, moreover, facilitate the discovery of super-fast rotating asteroids 
\citep[cf.][and references therein]{chang2017}.
Wide-field facilities such as ZTF are particularly powerful for this type of science because of the efficiency of collecting numerous light curves within a short period of time 
\citep[e.g.,][]{Masiero2009, Polishook2009, Dermawan2011, Polishook2012, Chang2014, Chang2015, Chang2016, Waszczak2015}. 
To demonstrate the ability of the ZTF for this task, we conducted a pilot campaign on December 15, 2017, in which we repeatedly scanned between two ZTF fields on the ecliptic plane at opposition for $\sim3$ hours using a cadence of 90 seconds. 
More than 2600 asteroid light curves with 10 or more detections were extracted by matching the source detections against the ephemerides obtained from the {\it JPL/HORIZONS} system with a search radius of 2\arcsec. 
To find the rotation periods of asteroids, we fitted all the light curves using a second-order Fourier series \citep{Harris1989}.
Due to the short observation time span, we were only able to detect periods of $< 3$ hour.
In Figure \ref{asteroid}a we show the ZTF lightcurve for asteroid (11014) Svatopluk folded to the derived rotation period of 2.25 hr.
However, most relatively bright asteroids show a clear light curve covering an incomplete rotation (Figure \ref{asteroid}b).
For faint asteroids ($\gtrsim 19.5$ mag), we were not able to conclusively identify any rotation periods (e.g., Figure \ref{asteroid}c), likely due to larger uncertainties masking the variability, and the short time span of observations. 
In this pilot campaign, we did not find any super-fast rotating asteroids. 

\subsubsection{Activity of Comets and Centaurs}
By covering the entire Northern sky approximately every three days (Sec.~\ref{sec:survey_strategy}), ZTF acquires serendipitous observations of a large number of comets and centaurs.  
Through ZTF's high cadence and sensitivity, it is well suited to monitor the activity development of comets and to look for temporal variability, including both secular changes and rotational modulation of the activity, as well as transient events such as outbursts. 

We identified comets and Centaurs in the ZTF data by comparing the telescope's observing logs to the ephemeris positions of all comets with predicted brightness $V<22$~mag. 
This brightness limit is below ZTF's detection limit,  but it is used not only because comet brightness predictions are notoriously poor, but also because an outburst could make a normally faint comet detectable. 
As of 24 April 2018, we estimate that ZTF had made 15000 observations of 186 comets brighter than 22~mag, and 3300 observations of 41 comets brighter than 18~mag.  

\begin{figure*}
\gridline{\fig{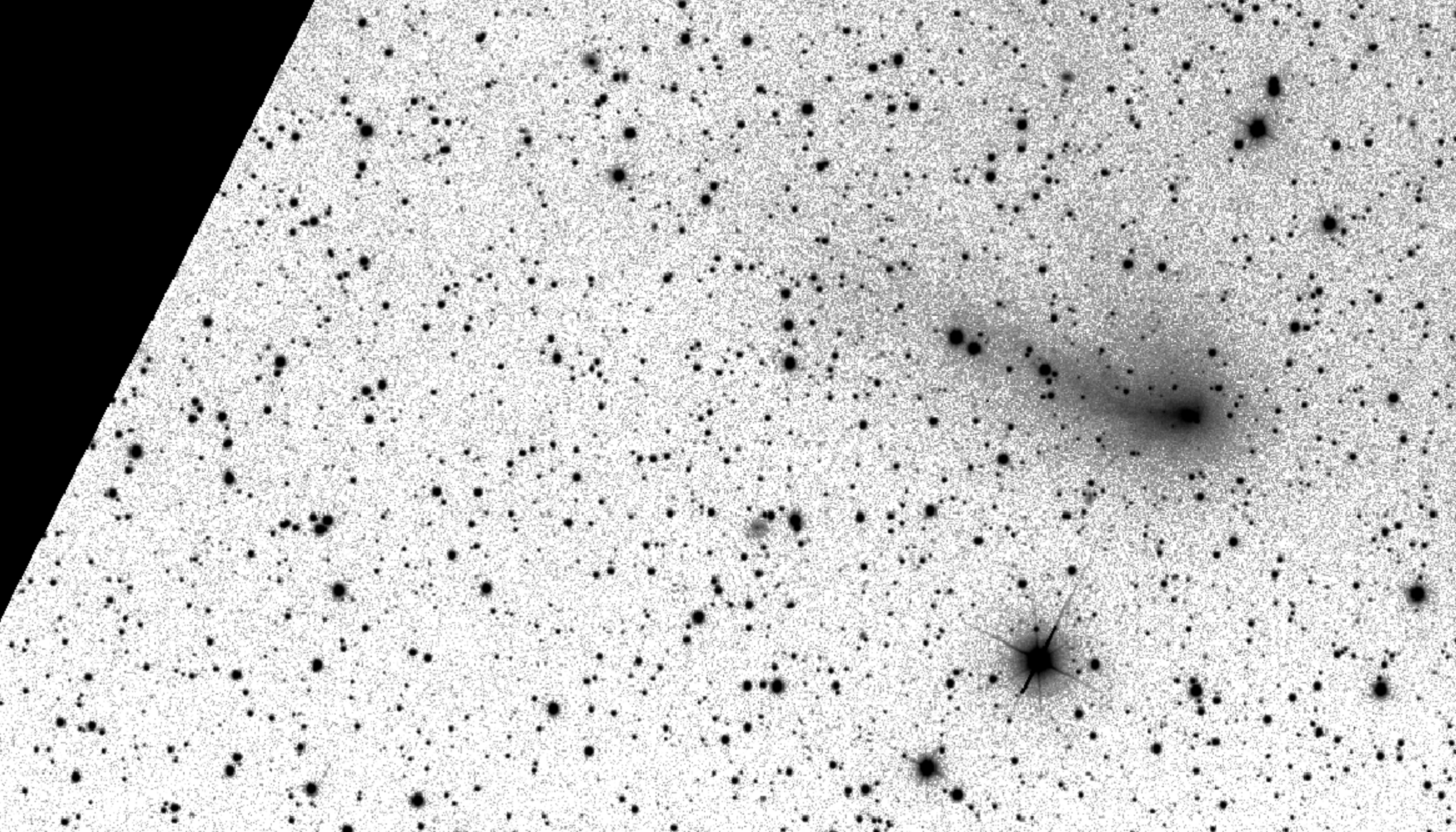}{0.5\textwidth}{(a)}
          \fig{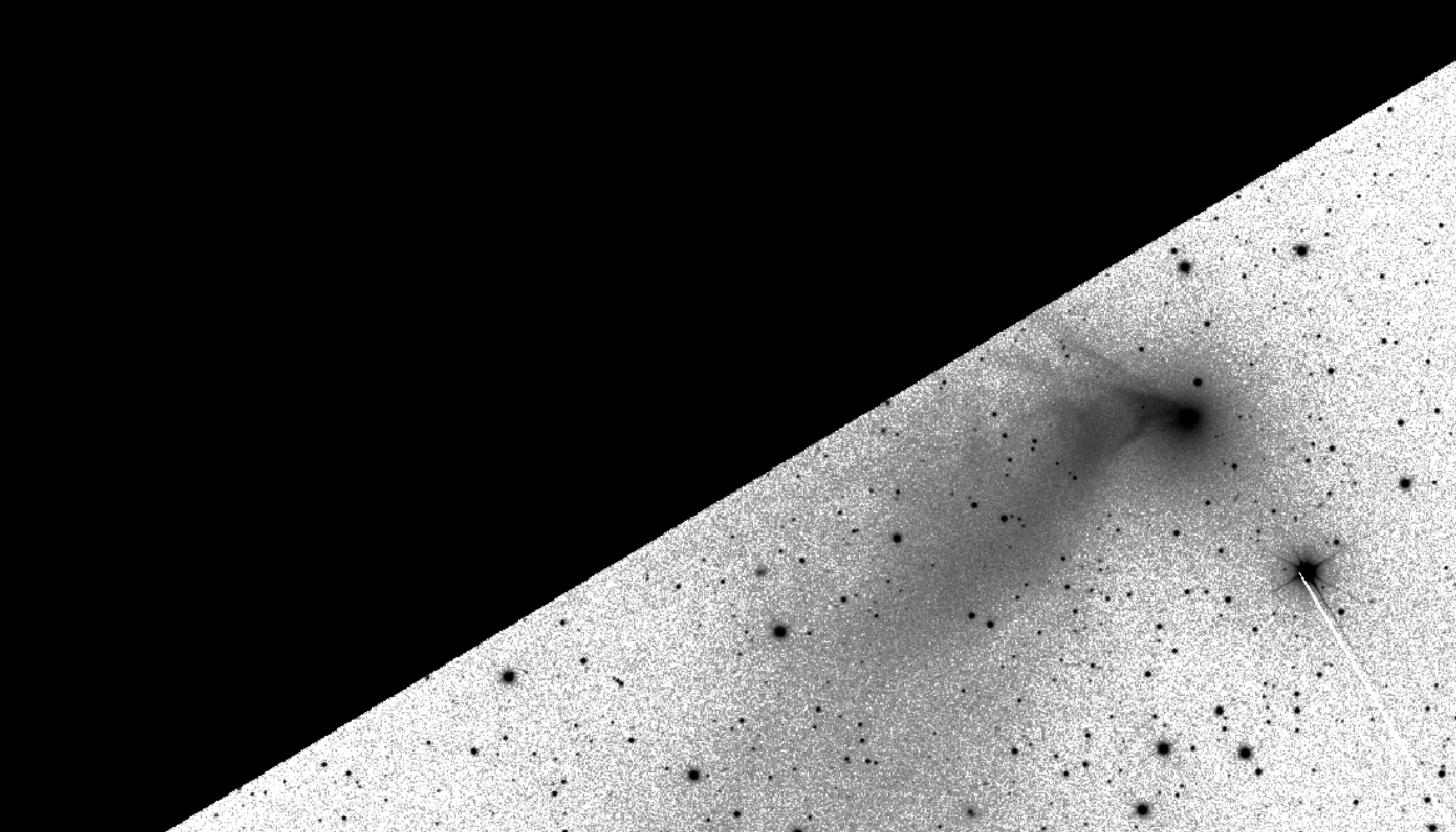}{0.5\textwidth}{(b)}}
\gridline{\fig{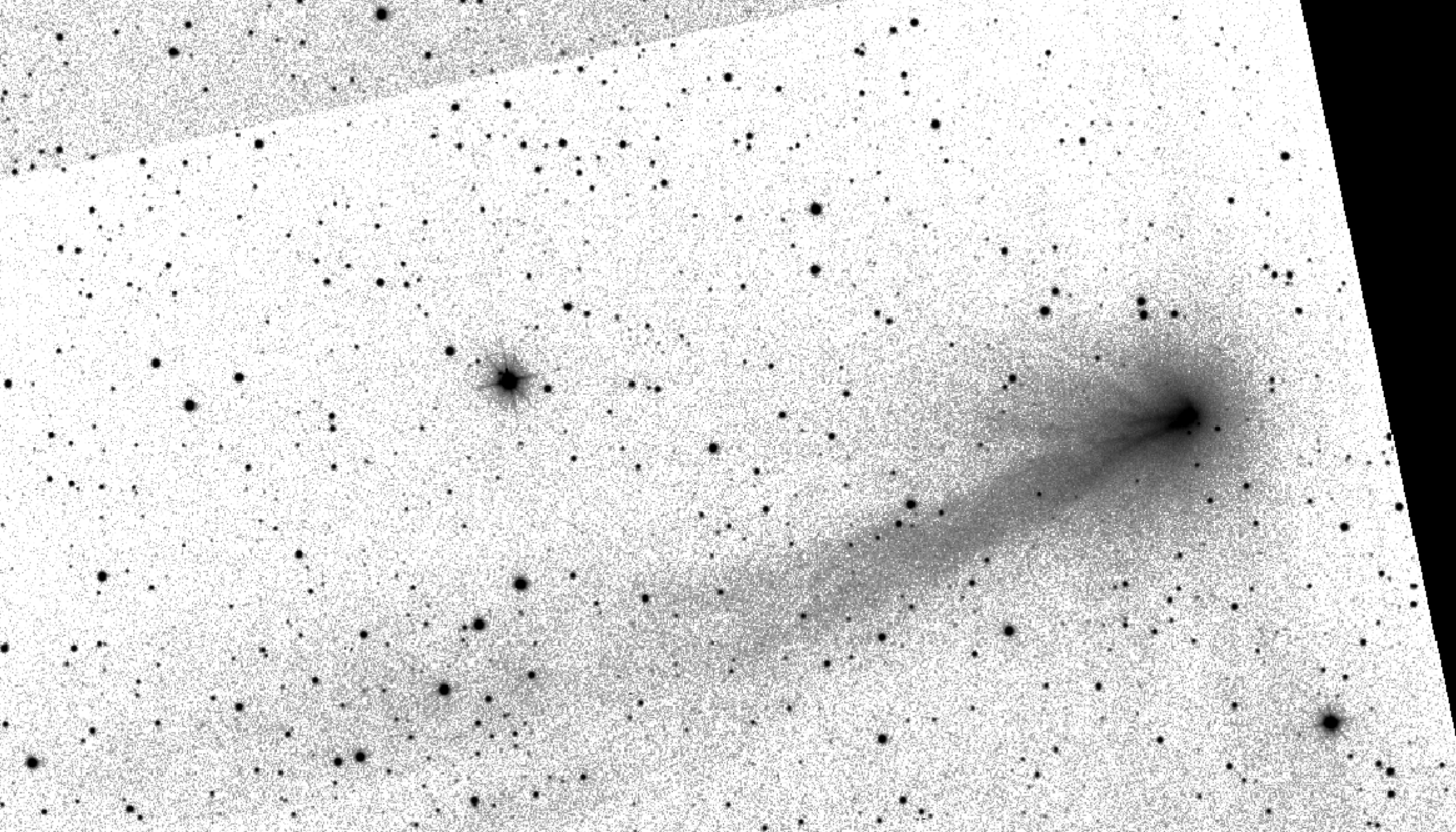}{0.5\textwidth}{(c)}
          \fig{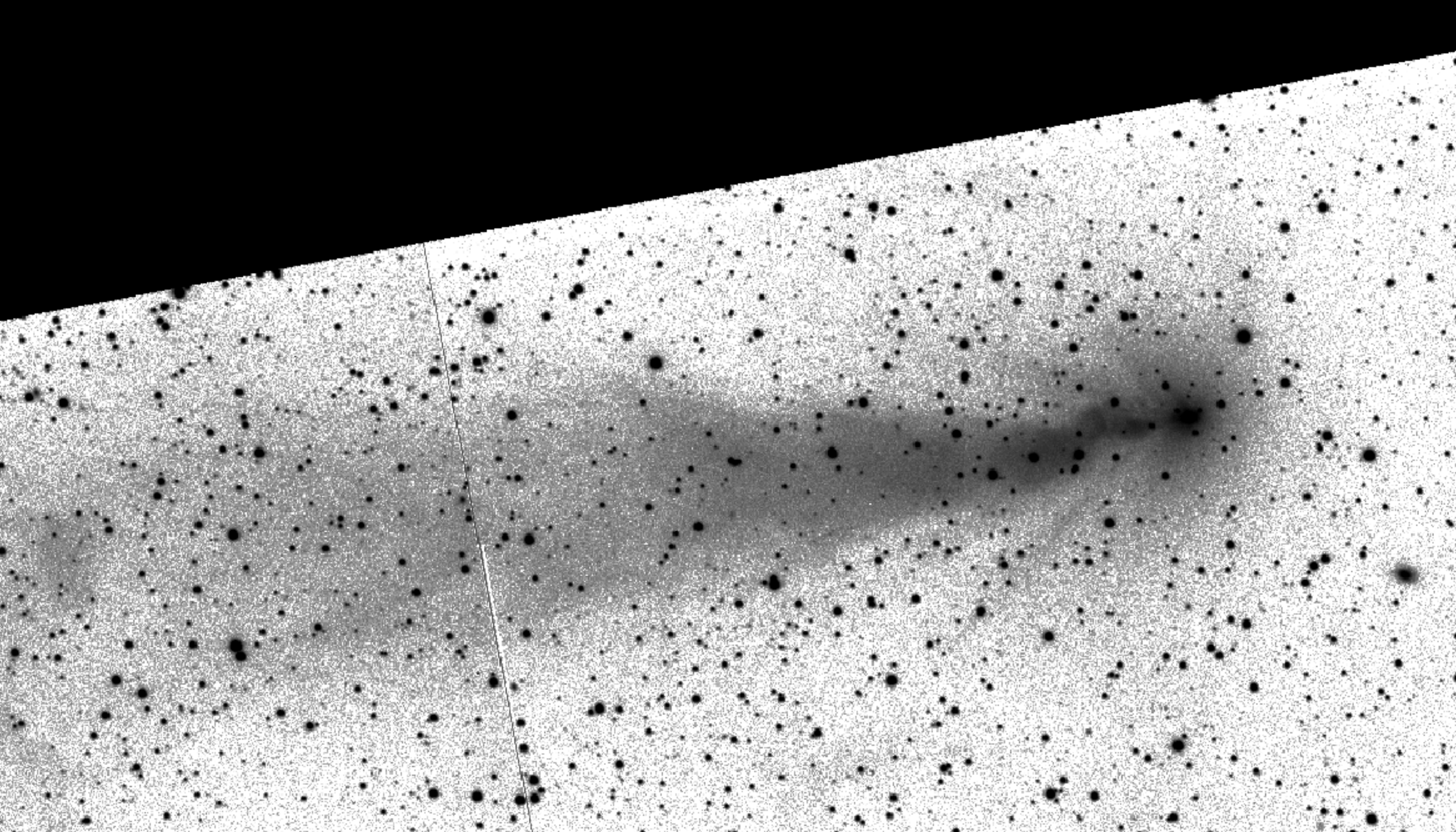}{0.5\textwidth}{(d)}}
\caption{ZTF images of comet C/2016 R2 (PanSTARRS) in the $g$-band.  Four epochs from the commissioning phase are shown: (a) 2017 Nov 11; (b) 2017 Dec 23; (c) 2018 Jan 13; and (d) 2018 Feb 19.  The field of view is $37\arcmin\times21\arcmin$ and the projected sunward vector is along the x-axis.  The images are logarithmically scaled, except near the background where they are linearly scaled, in order to enhance details in the tail.  
}
\label{comet}
\end{figure*}

ZTF imaging of C/2016~R2 (PanSTARRS) acquired between November 11, 2017 and February 19, 2018 is presented in Fig.~\ref{comet}. 
The images show the comet before perihelion, approaching the Sun from 3.2 to 2.7\,AU. 
At such heliocentric distances, water sublimation rates are low, yet the comet had an impressive ion tail spanning over $0.5^\circ$. 
This emission is fluorescence by CO$^+$ ions within the $g$ band \citep{Cochran2018}. 
No other volatiles have been detected and this comet appears to have an extremely high chemical abundance of CO \citep{Cochran2018}, suggesting that CO sublimation drives the activity of this comet. 
Changes in the morphology of the ion tail reflect temporal variations in the comet's activity and in the local solar wind conditions \citep[cf.][]{Jones2018}. 
ZTF monitoring will allow us to follow the comet's activity evolution until it falls below V > 21, anticipated around 8\,AU from the Sun (JPL/Horizons).  

\begin{figure*}
\begin{center}
\includegraphics[width=0.75\textwidth]{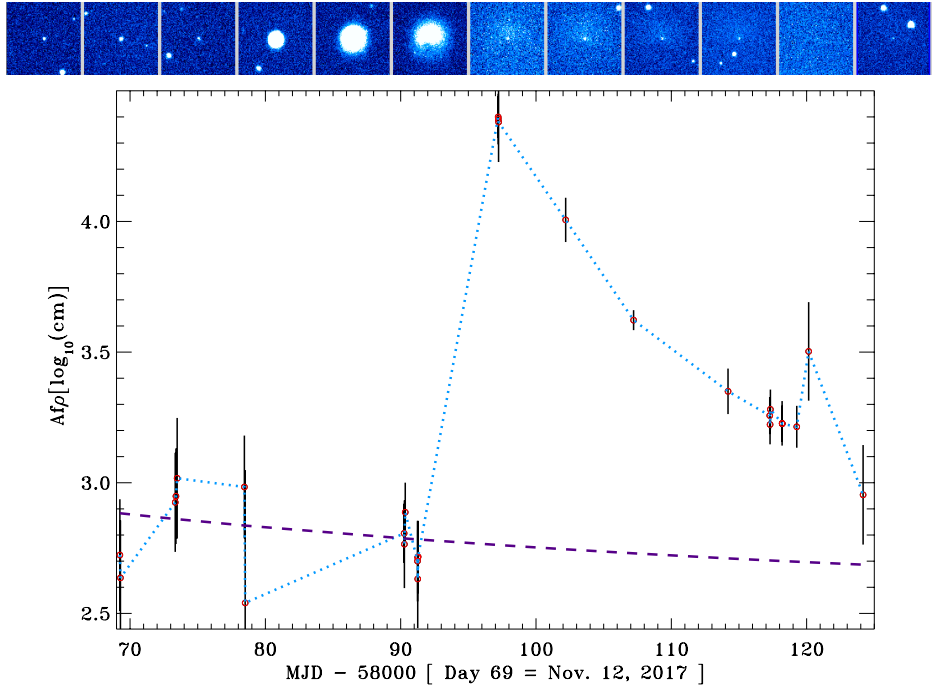}
\caption{ZTF observations of the outburst of comet Echeclus. The top panel shows the 1.5 arcmin FOV images of Echeclus, oriented North up and East to the left, on 12 Nov. 2017, 16 Nov. 2017, 04 Dec. 2017, 10 Dec. 2017, 15 Dec. 2017, 20 Dec. 2017, 27 Dec. 2017,  30 Dec. 2017, 31 Dec. 2017, 01 Jan. 2018, 02 Jan. 2018, and 06 Jan. 2018 (UT dates) left to right. The plot shows results from aperture photometry (7 arcsecond radius aperture) from the Echeclus data spanning dates between 12 November, 2017 through 16 January, 2018. We converted these to the equivalent Af$\rho$ values (a proxy for dust production) in log-cm units for the corresponding dates. The magenta dashed line indicates the derived Af$\rho$ value for a magnitude value corresponding to a bare nucleus. The images were primarily taken in the ZTF $r$-band, while those taken on the 12-16 Nov. 2017, or on or after 27 Dec. 2017 were ZTF $g$-band images.}
\label{Echec}
\end{center}
\end{figure*}

The first outburst observed by ZTF was seen when the Centaur Echeclus (q=5.8 AU, e=0.46, i=$4.3^\circ$) exhibited an increase in activity at 7.3 AU from the Sun. 
The outburst, originally discovered by Brian Skiff at Lowell Observatory, occurred on 7 December 2017 UT, and was first observed by ZTF on 10~December. It produced a dust coma with peak Af$\rho$ \citep[a proxy for dust production;][]{AHearn1984} of $20,000 \pm 2500$~cm (Fig.~\ref{Echec}), similar to previously observed outbursts of this object \citep{Bauer2008}. Assuming a dust ejection velocity near 50~m/s for $\sim 1\mu$m grain radii, we find a dust production rate $\sim$300~kg/s \citep[cf.][]{Bauer2008}. 
The August/September 2016 outburst produced brightening that lasted just over a month, while the late 2017 outburst also lasted roughly 30 days, as shown in the ZTF data.

\section{Summary}

ZTF will survey the Northern Hemisphere sky hundreds of times in three bands, with observations taken on timescales from minutes to years.
We expect the resulting datasets to enable discovery of young supernovae and rare relativistic transients; construction of systematic samples of Tidal Disruption Events, Active Galactic Nuclei, and variable stars; and detailed measurements of a variety of solar system objects.  
Thanks to ZTF's moderate depth, ZTF discoveries will be readily amenable to follow-up observations with 1--5\,m-class telescopes.
\citet{tmp_Graham:18:ZTFScience} provides a more thorough overview of the ZTF science case.

With the P48 focal plane now filled with CCDs, future sky surveys with the P48 will require substantial effort to achieve further performance improvements relative to ZTF, although further gains in angular resolution, wavelength coverage, and/or time sampling may be contemplated.  
Instead, most third-generation sky surveys will look to naturally scalable networks of small and medium telescopes distributed geographically, following the example of ASAS-SN, Las Cumbres Observatory, ATLAS, KMTNet, and BlackGEM.
The alternative model is large new monolithic facilities purpose-built for time-domain observations, with LSST serving as the exemplar.

Indeed, while the survey characteristics are quite different, ZTF will serve as a useful precursor for LSST.
ZTF will stream one million time-domain detections nightly using a prototype of the LSST alert distribution system, providing several years of community experience ahead of LSST's flood of ten million nightly alerts.

Because of its larger field of view, ZTF obtains on average about four times more observations of any area of the sky than LSST, and these visits are split among a smaller set of filters.
The resulting finer time-sampling will enable earlier discovery of transients and better classification of events based on their lightcurves.
Moreover, ZTF's smaller aperture means that all of the ZTF-discovered events are accessible for spectroscopic followup with moderate-aperture telescopes.
In fact, ZTF's discovery rate of transients brighter than 21$^\textrm{st}$ magnitude is greater than LSST's \citep{Bellm:16:Cadences}.
ZTF should thus provide large samples of bright transients and variables that will be crucial for interpreting LSST's deeper and more challenging survey.

\acknowledgments
Based on observations obtained with the Samuel Oschin Telescope 48-inch and the 60-inch Telescope at the Palomar Observatory as part of the Zwicky Transient Facility project. Major funding has been provided by the U.S.\ National Science Foundation under Grant No.\ AST-1440341 and by the ZTF partner institutions: the California Institute of Technology, the Oskar Klein Centre, the Weizmann Institute of Science, the University of Maryland, the University of Washington, Deutsches Elektronen-Synchrotron, the University of Wisconsin-Milwaukee, and the TANGO Program of the University System of Taiwan.

This work is partly based on observations made with the Nordic Optical Telescope, operated by the Nordic Optical Telescope Scientific Association at the Observatorio del Roque de los Muchachos, La Palma, Spain, of the Instituto de Astrofisica de Canarias. The data presented here were obtained in part with ALFOSC, which is provided by the Instituto de Astrofisica de Andalucia (IAA) under a joint agreement with the University of Copenhagen and NOTSA. This work is partly based on observations made with DOLoRes@TNG.

J.~Bauer, T.~Farnham, and M.~Kelley gratefully acknowledge the NASA/University of Maryland/MPC Augmentation through the NASA Planetary Data System Cooperative Agreement NNX16AB16A.

E.\ Bellm, B.\ Bolin, A.\ Connolly, V.~Z.\ Golkhou, D.\ Huppenkothen, Z.\ Ivezi\'{c}, L.\ Jones, M.\ Juric, and M.\ Patterson 
acknowledge support from the University of Washington College of Arts and Sciences, Department of Astronomy, and the DIRAC Institute. University of Washington's DIRAC Institute is supported through generous gifts from the Charles and Lisa Simonyi Fund for Arts and Sciences, and the Washington Research Foundation. M.~Juric and A.~Connolly acknowledge the support of the Washington Research Foundation Data Science Term Chair fund, and the UW Provost's Initiative in Data-Intensive Discovery.

E.\ Bellm, A.\ Connolly, Z.\ Ivezi\'{c}, L.\ Jones, M.\ Juric, and M.\ Patterson 
acknowledge support from the Large Synoptic Survey Telescope, which is supported in part by the National Science Foundation through
Cooperative Agreement 1258333 managed by the Association of Universities for Research in Astronomy
(AURA), and the Department of Energy under Contract No. DE-AC02-76SF00515 with the SLAC National
Accelerator Laboratory. Additional LSST funding comes from private donations, grants to universities,
and in-kind support from LSSTC Institutional Members.

E.~Bellm is supported in part by the NSF AAG grant 1812779 and grant \#2018-0908 from the Heising-Simons Foundation.

B.T. Bolin acknowledges funding for the Asteroid Institute program provided by B612 Foundation, W.K. Bowes Jr. Foundation, P. Rawls Family Fund and two anonymous donors in addition to general support from the B612 Founding Circle.

M. Bulla acknowledges support from the Swedish Research Council (Vetenskapsr\aa det) and the Swedish National Space Board.

C.-K.~Chang, W.-H.~Ip, C.-D.~Lee, Z.-Y.~Lin, C.-C.~Ngeow and P.-C.~Yu thank the funding from Ministry of Science and Technology (Taiwan) under grant 104-2923-M-008-004-MY5, 104-2112-M-008-014-MY3, 105-2112-M-008-002-MY3, 106-2811-M-008-081 and 106-2112-M-008-007.

A.\ Gal-Yam is supported by the EU via ERC grant No. 725161, the Quantum Universe I-Core program, the ISF, the BSF Transformative program and by a Kimmel award. 

S.~Gezari is supported in part by NSF CAREER grant 1454816 and NSF AAG grant 1616566.

D.~A.~Goldstein acknowledges support from Hubble Fellowship grant HST-HF2-51408.001-A. Support for Program number HST-HF2-51408.001-A is provided by NASA through a grant from the Space Telescope Science Institute, which is operated by the Association of Universities for Research in Astronomy, Incorporated, under NASA contract NAS5-26555.

M.~M.\ Kasliwal and Q.-Z.\ Ye acknowledge support by the GROWTH (Global Relay of Observatories Watching Transients Happen) project funded by the National Science Foundation PIRE (Partnership in International Research and Education) program under Grant No 1545949.

A.~A.\ Mahabal acknowledges support from the following grants: NSF AST-1749235, NSF-1640818 and NASA 16-ADAP16-0232.

A.~A.\ Miller is funded by the Large Synoptic Survey Telescope Corporation in support of
the Data Science Fellowship Program.

E.~Ofek is grateful for support by  a grant from the Israeli Ministry of Science,  ISF, Minerva, BSF, BSF transformative program, and  the I-CORE Program of the Planning  and Budgeting Committee and The Israel Science Foundation (grant No 1829/12).

M.~Rigault is supported by the European Research Council (ERC) under the European Union's Horizon 2020 research and innovation programme (grant agreement no.\ 759194 - USNAC).

J.~Sollerman acknowledges support from the Knut and Alice Wallenberg Foundation.

M.~T.\ Soumagnac acknowledges support by a grant from IMOS/ISA, the Ilan Ramon fellowship from the Israel Ministry of Science and Technology and the Benoziyo center for Astrophysics at the Weizmann Institute of Science.

Part of this research was carried out at the Jet Propulsion Laboratory, California Institute of Technology, under a contract with the National Aeronautics and Space Administration.

\facilities{PO:1.2m, PO:1.5m, ARC, DCT, NOT}


\appendix

\section{Observations of ZTF18aaayemw}
\label{sec:SNdetails}

\subsection{Light curves}
Host-subtracted PSF photometry of ZTF18aaayemw was produced from our P48 observations using a Pan-STARRS1 $r$-band stack as the reference image, since our P48 references contain SN light. The image subtraction and photometry methods follow \citet{fst+16}.

In addition to the P48 observations, we observed ZTF18aaayemw with the Centurion 28-inch telescope (C28) and 1-m telescope at the WISE observatory (Israel). 
In addition, we also obtained several epochs of UV photometry with Ultraviolet/Optical Telescope \citep[UVOT;][]{Roming2005a} of the Neil Gehrels \textit{Swift} Observatory \citep{Gehrels2004a}. 
The data were reduced with routines in IRAF \citep{Tody1986a} version 2.16. 
The world-coordinate system was calibrated with the software package \texttt{astrometry.net} \citep{Lang2010a} version 0.69. 
We measured the brightness of the transients using circular apertures in Source Extractor version 2.19.5 \citep{Bertin1996a} where the aperture diameter had a size of $1\times {\rm FWHM}$ of the stellar point-spread function. 
The absolute flux calibration was secured with the Sloan Digital Sky Survey DR12 \citep{Alam2015a}. The UVOT data were retrieved from the \textit{Swift} Data Archive\footnote{\url{http://www.swift.ac.uk/swift\_portal/}}. 
We used the standard UVOT data analysis software distributed with HEASOFT version 6.19, along with the standard calibration data.
All photometry is summarized in Table~\ref{tab:phot}, and has not been corrected for foreground or host reddening.

\subsection{Spectroscopy}
Spectra of ZTF18aaayemw were obtained with the AndaLucia Faint Object Spectrograph and Camera (ALFOSC) on the Nordic Optical Telescope, with the Double-Beamed Spectrograph (DBSP; \citealt{og82}) on the 200-in Hale Telescope at Palomar Observatory, the Device Optimized for the LOw RESolution (DOLORES) on the Telescopio Nazionale Galileo, the DeVeny Spectrograph on the Discovery Channel Telescope, and the Dual Imager Spectrograph on the 3.5 ARC telescope at Apache Point Observatory. 
Details of the observations are listed in Table~\ref{tab:spec}.


NOT and TNG spectra were reduced using a combination of IRAF and MATLAB scripts, which included bias and flat-field corrections; extraction of the 1D spectrum; wavelength calibration of the spectrum by comparison with the spectrum of an arc lamp; flux calibrations using the sensitivity function built from the spectra of a spectral standard star observed the same night. The TNG spectra from the two different grisms (see Table~\ref{tab:spec}) were combined together.

The APO+DIS spectrum was reduced using {\tt pydis}\footnote{\url{https://github.com/TheAstroFactory/pydis}}. A spectrophotometric standard star observed on the same night in the same instrumental configuration was used for flux calibration.

The DCT DeVeny spectrum was reduced using standard \texttt{IRAF} routines. We first corrected for bias and flat-field then extracted the 1D spectrum. Wavelength and flux calibration were done by using a comparing with spectra of an arc lamp and the flux standard Feige34.


\bigskip
\begin{deluxetable}{lccccc}
\tablecaption{Log of ZTF18aaayemw Spectroscopic Observations
\label{tab:spec}}
\tablehead{
\colhead{Observation Date (UT)} &
\colhead{Telescope+Instrument} &
\colhead{Slit} &
\colhead{Grating} &
\colhead{Exp. Time (s)} &
\colhead{Airmass}
}
\startdata
2018-02-14 & NOT+ALFOSC & 1.0" & gr\#4 & 2400   & 1.08 \\
2018-02-19 & P200+DBSP   & 1.5"& 600/4000 & 1200 & 1.21 \\
2018-02-21 & TNG+DOLORES & 1.5"& LR-B+LR-R & 1800+1500 & 1.09 \\
2018-02-22 & DCT+LMI & 1.5"& 300g/mm & 8100 & 1.08 \\
2018-03-20 & APO+DIS & 1.5" & B400/R300 & 3600 & 1.0 \\
2018-03-23 & NOT+ALFOSC & 1.0" & gr\#4 & 4800 & 1.05\\
\enddata
\end{deluxetable}

\startlongtable
\begin{deluxetable}{lccc}
\tablecaption{ZTF18aaayemw Light Curve
\label{tab:phot}}
\tablehead{
\colhead{MJD} &
\colhead{Filter} &
\colhead{AB mag} &
\colhead{Instrument}
}
\startdata
58154.26 &    R & $19.95 \pm 0.08$ & P48+ZTF \\
58154.26 &    R & $20.10 \pm 0.10$ & P48+ZTF \\
58154.28 &    R & $20.22 \pm 0.14$ & P48+ZTF \\
58154.30 &    R & $20.12 \pm 0.11$ & P48+ZTF \\
58155.24 &    R & $19.66 \pm 0.06$ & P48+ZTF \\
58155.24 &    R & $19.61 \pm 0.05$ & P48+ZTF \\
58155.26 &    R & $19.45 \pm 0.05$ & P48+ZTF \\
58155.26 &    R & $19.60 \pm 0.05$ & P48+ZTF \\
58155.28 &    R & $19.57 \pm 0.04$ & P48+ZTF \\
58155.30 &    R & $19.58 \pm 0.06$ & P48+ZTF \\
58155.32 &    R & $19.64 \pm 0.05$ & P48+ZTF \\
58156.24 &    R & $19.36 \pm 0.05$ & P48+ZTF \\
58156.26 &    R & $19.36 \pm 0.06$ & P48+ZTF \\
58156.28 &    R & $19.38 \pm 0.06$ & P48+ZTF \\
58156.31 &    R & $19.34 \pm 0.06$ & P48+ZTF \\
58156.33 &    R & $19.22 \pm 0.03$ & P48+ZTF \\
58158.25 &    R & $18.89 \pm 0.02$ & P48+ZTF \\
58158.25 &    R & $18.96 \pm 0.03$ & P48+ZTF \\
58158.26 &    R & $18.83 \pm 0.03$ & P48+ZTF \\
58158.26 &    R & $18.91 \pm 0.02$ & P48+ZTF \\
58158.28 &    R & $18.97 \pm 0.03$ & P48+ZTF \\
58158.31 &    R & $18.96 \pm 0.03$ & P48+ZTF \\
58158.31 &    R & $18.94 \pm 0.02$ & P48+ZTF \\
58158.32 &    R & $18.96 \pm 0.02$ & P48+ZTF \\
58158.32 &    R & $18.89 \pm 0.03$ & P48+ZTF \\
58160.28 &    R & $18.73 \pm 0.03$ & P48+ZTF \\
58160.30 &    R & $18.61 \pm 0.03$ & P48+ZTF \\
58160.30 &    R & $18.65 \pm 0.04$ & P48+ZTF \\
58160.32 &    R & $18.71 \pm 0.03$ & P48+ZTF \\
58160.32 &    R & $18.58 \pm 0.03$ & P48+ZTF \\
58160.34 &    R & $18.71 \pm 0.03$ & P48+ZTF \\
58160.34 &    R & $18.67 \pm 0.03$ & P48+ZTF \\
58183.25 &    R & $18.83 \pm 0.07$ & P48+ZTF \\
58183.25 &    R & $18.92 \pm 0.07$ & P48+ZTF \\
58183.27 &    R & $18.91 \pm 0.07$ & P48+ZTF \\
58184.24 &    R & $18.97 \pm 0.03$ & P48+ZTF \\
58184.24 &    R & $18.93 \pm 0.03$ & P48+ZTF \\
58184.26 &    R & $18.95 \pm 0.03$ & P48+ZTF \\
58184.26 &    R & $18.94 \pm 0.03$ & P48+ZTF \\
58184.28 &    R & $18.94 \pm 0.04$ & P48+ZTF \\
58184.28 &    R & $18.97 \pm 0.03$ & P48+ZTF \\
58186.27 &    R & $18.95 \pm 0.02$ & P48+ZTF \\
58186.29 &    R & $18.96 \pm 0.02$ & P48+ZTF \\
58175.58 &    U & $19.45 \pm 0.13$ & UVOT \\
58177.64 &    U & $19.68 \pm 0.15$ & UVOT \\
58178.05 &    U & $19.92 \pm 0.15$ & UVOT \\
58175.57 & UVM2 & $20.69 \pm 0.18$ & UVOT \\
58177.63 & UVM2 & $21.54 \pm 0.30$ & UVOT \\
58178.04 & UVM2 & $21.26 \pm 0.23$ & UVOT \\
58175.58 & UVW1 & $20.31 \pm 0.16$ & UVOT \\
58177.64 & UVW1 & $21.28 \pm 0.30$ & UVOT \\
58178.04 & UVW1 & $20.61 \pm 0.18$ & UVOT \\
58175.58 & UVW2 & $21.08 \pm 0.17$ & UVOT \\
58177.65 & UVW2 & $21.39 \pm 0.20$ & UVOT \\
58178.05 & UVW2 & $21.35 \pm 0.19$ & UVOT \\
58172.82 &    R & $18.40 \pm 0.04$ & C28 \\
58172.83 &    R & $18.34 \pm 0.04$ & C28 \\
58172.83 &    R & $18.37 \pm 0.04$ & C28 \\
58188.71 &    R & $18.95 \pm 0.04$ & WISE-1m \\
58188.72 &    R & $18.95 \pm 0.03$ & WISE-1m \\
58188.73 &    R & $18.94 \pm 0.03$ & WISE-1m \\
58189.70 &    R & $18.98 \pm 0.04$ & WISE-1m \\
58189.71 &    R & $19.01 \pm 0.05$ & WISE-1m \\
58189.78 &    R & $18.95 \pm 0.04$ & WISE-1m \\
58189.80 &    R & $18.94 \pm 0.03$ & WISE-1m \\
58189.72 &    u & $21.17 \pm 0.32$ & WISE-1m \\
\enddata
\end{deluxetable}

\end{CJK*}
\end{document}